\documentclass[journal,twoside]{IEEEtran}
\usepackage{amsmath,amssymb,url}
\newtheorem{theorem}{Theorem}
\newtheorem{lemma}{Lemma}
\newtheorem{defn}{Definition}
\newtheorem{prpn}{Proposition}

\newtheorem{example}{Example}

\ifCLASSINFOpdf
\else
\fi
\usepackage{array}

\begin{document}
\title{Low-delay, Low-PAPR, High-rate Non-square Complex Orthogonal Designs}

\author{Smarajit Das,~\IEEEmembership{Student Member,~IEEE} and
        B. Sundar Rajan,~\IEEEmembership{Senior Member,~IEEE}%
\thanks{This work was supported through grants to B.S.~Rajan; partly by the
IISc-DRDO program on Advanced Research in Mathematical Engineering, and partly
by the Council of Scientific \& Industrial Research (CSIR, India) Research
Grant (22(0365)/04/EMR-II). Part of the material in this paper was presented in IEEE 2008 International Symposium on Information Theory (ISIT-2008), July 6-11, Toronto, Canada.} 
\thanks{Smarajit Das and B. Sundar Rajan are with the Department of Electrical Communication Engineering, Indian Institute of Science, Bangalore-560012, India.email:\{smarajit,bsrajan\}@ece.iisc.ernet.in.}}

\maketitle
\begin{abstract}
The maximal rate for non-square Complex Orthogonal Designs (CODs) with $n$ transmit antennas is  $\frac{1}{2}+\frac{1}{n}$ if $n$ is even  and  $\frac{1}{2}+\frac{1}{n+1}$ if $n$ is odd, which are close to $\frac{1}{2}$ for large values of $n.$  A class of maximal rate non-square CODs have been constructed by Liang (IEEE Trans. Inform. Theory, 2003) and  Lu et. al. (IEEE Trans. Inform. Theory, 2005) have shown that the decoding delay of the codes given by Liang, can be reduced by $50\%$ when number of transmit antennas is a multiple of $4$. Adams et. al. (IEEE Trans. Inform. Theory, 2007) have shown that the designs of Liang are of minimal-delay for $n$ equal to $1$ and $3$ modulo 4 and that of Lu et.al. are of minimal delay when $n$ is a multiple of $4.$ However, these minimal delays are large compared to the delays of the rate $1/2$ non-square CODs constructed by Tarokh et al (IEEE Trans. Inform. Theory, 1999) from rate-$1$ real orthogonal designs (RODs). In this paper, we construct a class of rate-$1/2$ non-square CODs for any $n$  with the decoding delay equal to  $50\%$ of that of the delay of the rate-1/2 codes given by Tarokh et al. This is achieved by giving first a general construction of rate-1 square Real Orthogonal Designs (RODs) which includes as special cases the well known constructions of Adams, Lax and Phillips and Geramita and Pullman, and then making use of it to obtain the desired rate-$\frac{1}{2}$ non-square COD. For the case of 9 transmit antennas, our rate-$\frac{1}{2}$ COD is shown to be of minimal-delay. The proposed construction results in designs with zero entries which may have high Peak-to-Average Power Ratio (PAPR) and it is shown that by appropriate postmultiplication, a design with no zero entries can be obtained with no change in the code parameters. 
\end{abstract}
\section{Introduction and Preliminaries}
\label{sec1}
There are several definitions of Orthogonal Designs (ODs) in the literature \cite{Lia,TJC,TWMS} the well known being as given in \cite{Lia}:
\begin{defn}
\label{cod} 
A  Complex Orthogonal Design (COD) $G(x_0,x_1,\cdots,x_{k-1})$  (in short $G$) for $n$ transmit antennas is defined as a $ p\times n$ matrix such that (i) the nonzero entries of $G$ are the complex variables $\pm x_0 , \pm x_1 , . . . , \pm x_{k-1}$ and their conjugates and (ii) $G^\mathcal{H} G=({\vert x_0\vert}^2 +\cdots+{\vert x_{k-1}\vert}^2)I_n$ where $\mathcal{H}$ stands for the complex conjugate transpose and $I_n$ is the $n\times n$ identity matrix. The matrix $G$ is also said to be a $[p,n,k]$ COD and its rate in complex symbols per channel use is $\frac{k}{p}.$ When $x_0,\cdots,x_{k-1}$ are real variables, the designs are called Real Orthogonal Design (ROD). 
\end{defn}

Space-time block codes (STBCs) from CODs have been widely studied for square designs, i.e., $p=n,$ since they correspond to minimum decoding delay codes for co-located multiple-antenna coherent communication systems. However, non-square designs naturally appear and important in the following  situations.

\begin{enumerate}
\item In coherent co-located MIMO systems, for a specified number of transmit antennas, non-square designs can give much higher rate than the square designs \cite{Lia}.
\item In non-coherent MIMO systems with non-differential detection, non-square designs with $p=2n$ lead to low decoding complexity STBCs \cite{TaK}.
\item Space-Time-Frequency codes can be viewed as non-square designs \cite{JSKR}.
\item In distributed space-time coding for relay channels, rectangular designs naturally appear \cite{JiH}.
\item Rate $\frac{1}{2}$ non-square CODs have been proposed for use in analog transmission with application to channel feedback \cite{ChS}.
\end{enumerate}

The rate of the square CODs falls exponentially with increase in the number of transmit antennas. The following theorem relates the rate of a square OD, real/complex, with the number of transmit antennas. For an integer $n=2^a(2b+1),$ where  $a=4c+d$ and $a,b,c$ and $d$ being integers with $ 0\leq d\leq 3,$ the  Hurwitz-Radon number $\rho (n)$ is defined as  $\rho(n)=8c+2^d.$

\begin{theorem}[~\cite{TiH},~\cite{TJC},~\cite{ALP}]
\label{thm1}
The maximal rate of a square ROD for $n$ transmit antennas is given by $\frac{\rho(n)}{n}$ where $\rho(n)$ is the Hurwitz-Radon number of $n$, while that of a square COD is given by $\frac{a+1}{n}$ where $a$ is the exponent of $2$ in the prime factorization of $n.$
\end{theorem}

Several authors have constructed square CODs achieving maximal rate \cite{TiH, ALP}. In \cite{TiH}, the following induction method is used to construct square CODs for $2^a$ antennas, $a=2,3,\cdots$, starting from 
\begin{equation}
\label{itcod}
G_1= \left[\begin{array}{rr}
x_0   &-x_1^*      \\
x_1   & x_0^*
\end{array}\right],~
G_a=  \left[\begin{array}{rr}
G_{a-1}   & -x_{a}^*I_{2^{a-1}}    \\
x_{a}I_{2^{a-1}}   & G_{a-1}^\mathcal{H}
\end{array}\right],
\end{equation}
\noindent
where $G_a$ is a $2^a\times 2^a$ complex matrix. Note that $G_a$ is a square COD in $(a+1)$ complex variables $x_0,x_1,x_2,\cdots, x_{a}$.

It is clear from Theorem \ref{thm1} as well as the construction given by \eqref{itcod} that the square ODs, real and complex are not bandwidth efficient and naturally one is led to study non-square ODs in order to obtain codes with higher rate. It is known that \cite{TJC} there always exists a rate-1 ROD for any number of transmit antennas. In fact, the existence of rate-1 $[p,n,p]$ ROD is equivalent to that of a $[p,p,n]$ square ROD. For a rate-1 ROD, the minimum value of decoding delay $p$ as a function of $n$, denoted by $\nu(n)$, is given by

{\footnotesize
\begin{eqnarray}
\label{nun}
\begin{array}{ll}
\nu(n)=2^{\delta(n)}\mbox{   where   }\\
 \delta(n)=
    \begin{cases}
4s      & \text{ if $n=8s+1$ } \\
4s+1    & \text{ if $n=8s+2$ } \\
4s+2    & \text{ if $n=8s+3$ or $8s+4$ }\\
4s+3    & \text{ if $n=8s+5,8s+6,8s+7$ or $8s+8$ }.
  \end{cases}
\end{array}
\end{eqnarray}
}
It is known \cite{Lia} that the maximal rate of  a non-square COD is equal to $\frac{1}{2}+\frac{1}{2a}$ when the number of transmit antennas is $2a-1$ or $2a$. Liang in \cite{Lia} has given an explicit construction of non-square CODs achieving this maximal rate for any number of antennas. There is also another construction of these codes given by Lu et al \cite{LFX}. The former construction is algorithmic in nature while the latter one is based on {\it patch-up} of several matrices. The minimum decoding delay for the maximal rate non-square CODs is, in general, not known. The following theorem states what is known about the minimum delay of these code. For details, see \cite{AKP}.
\begin{theorem}[~\cite{AKP}]
\label{delaycod}
A tight lower bound on the decoding delay of maximum rate non-square CODs for $n$ antennas is $\binom{2a}{a-1}$ for $n=2a-1$ or $n=2a$. Moreover, if $n$ is congruent to $0,1$ or $3$ modulo $4$, then this lower bound is achievable. If $n$ is congruent to $2$ modulo $4$, the minimum decoding delay is upper bounded by $2 \binom{2a}{a-1}$.
\end{theorem}
As the rate of these maximal rate codes is close to $1/2$ for large number of antennas, it is sufficient to focus on rate $1/2$ codes when large number of antennas is under consideration. Then, a natural problem to study is construction of rate $1/2$ non-square CODs with the decoding delay as small as possible. 
Tarokh et al \cite{TJC} have given a class of rate $1/2$ codes obtained from rate-1 RODs, which has lower delays than those of maximal-rate codes constructed in \cite{Lia,LFX} for number of transmit antennas more than 5. For example, for four transmit antennas, the rate $1/2$ code is
\begin{equation}  
G=\frac{1}{\sqrt{2}}\left[\begin{array}{rrrrrrrr}
    x_0   &-x_1   &-x_2  &-x_3   \\
    x_1   & x_0   & x_3  &-x_2   \\
    x_2   &-x_3   & x_0  & x_1  \\
    x_3   & x_2   &-x_1  & x_0  \\
    x_0^* &-x_1^*   &-x_2^*  &-x_3^*   \\
    x_1^*   & x_0^*   & x_3^*  &-x_2^*   \\
    x_2^*   &-x_3^*   & x_0^*  & x_1^*  \\
    x_3^*   & x_2^*   &-x_1^*  & x_0^*  
    \end{array}\right].
\end{equation}

Notice that, contrary to the definition in \cite{TJC} of a COD,  in this code, the variables appear in a column more than once and each entry of all the columns of the design matrix is scaled by $\frac{1}{\sqrt{2}}$ in order to satisfy the condition $G^\mathcal{H} G=({\vert x_0\vert}^2 +\cdots+{\vert x_3\vert}^2)I_4$ of a COD. We call such designs {\it scaled} COD which is not a COD in the conventional sense as in \cite{Lia} (Definition \ref{cod}). We define the class of scaled CODs as follows:
\begin{defn}
\label{scod}
A {\textit {$\lambda$-scaled complex orthogonal design}}, for a positive integer $\lambda,$ ($\lambda$-scaled COD) $G$ is a $p\times n$ matrix in $k$ complex variables $x_0,x_1,\cdots,x_{k-1}$ such that a non-zero entry of the matrix is a variable or its complex conjugate, or the negative of these and all the entries of any subset of  columns of the matrix is scaled by $\frac{1}{\sqrt{\lambda}}$ satisfying the condition: $G^\mathcal{H} G=({\vert x_0\vert}^2 +\cdots+{\vert x_{k-1}\vert}^2)I_n.$  The matrix $G$ is also said to be a $[p,n,k]$ $\lambda-$scaled COD. 
\end{defn}

Notice that a $\lambda$-scaled COD with with no column scaled by $\frac{1}{\sqrt{\lambda}}$ is a COD. In columns with scaling by $\frac{1}{\sqrt{\lambda}}$ all the variables appear exactly $\lambda$ times. In this paper we consider only the case $\lambda=2$ and call these codes simply scaled-COD. 

{\it Contributions of this paper:} The contributions of this paper may be summarized as follows: 
\begin{itemize} 
\item For the rate  $1/2$ scaled CODs of \cite{TJC}, all the columns are scaled by the factor $\frac{1}{\sqrt{2}},$ which led to the reduced delay compared to the codes of Liang and a main result of this  paper is that by having only a subset of the columns scaled by $\frac{1}{\sqrt{2}}$ further reduction in delay by 50\% is possible. We use following notations to refer to the rate 1/2 CODs given by Tarokh et al \cite{TJC}, the maximal rate codes given in Lu et al \cite{LFX} and the codes of this paper.
\begin{itemize}
\item
$L_n$ is the maximal rate COD for $n$ transmit antennas with the decoding delay as specified in the Theorem~\ref{delaycod}. 
\item
$TJC_n$ is the rate $1/2$ scaled CODs for $n$ transmit antennas constructed by Tarokh et al \cite{TJC}.
\item $(DR)_n$ is the rate $1/2$ scaled CODs for $n$ transmit antennas constructed in this paper.
\end{itemize}
Note that as $n$ increases, the maximal rate of $L_n$ approaches $1/2$, thus two codes $L_n$ and $TJC_n$ can be compared for large value of $n,$ based on their delays. It is not difficult to see that the decoding delay of $TJC_n$ is less than that of $L_n$ for large $n$. We provide an explicit construction of rate $1/2$ scaled CODs for any number of transmit antennas such that the decoding delay of these codes is $\nu(n)$ when $n$ is the number of transmit antennas, whereas the delay for the codes $TJC_n$ is $2\nu(n).$ The Table \ref{recparameters} at the top of the next page shows that for large values of $n,$ but for a marginal decrease in the rate with respect to $L_n,$ the codes of this paper are the best codes.
\item For the case of 9 transmit antennas our rate-$\frac{1}{2}$ code is shown to be of minimal-delay.

\begin{table*}
\caption{The comparison of maximum rate achieving codes and rate 1/2  codes}
\label{recparameters}
\begin{center}
\begin{tabular}{c||rrrrrrrrrrrrrrrr}
$n$                     &5 &6 &7    &8  &9 &10  &11 &12 &13 &14 &15 &16 \\ \hline \hline
Decoding delay of $(DR)_n$ &8 &8 &8 &8 &16 &32 &64 &64 &128 &128 &128 &128 \\ \hline
Decoding delay of $TJC_n$ &16 &16 &16 &16 &32 &64 &128 &128 &256 &256 &256 &256 \\ \hline
\mbox{ Decoding delay of } $L_n$ &15 &30 &56 &56 &210 &420 &792 &792  &3003 & 6006& 11440& 11440\\ \hline \hline
$\mbox{ Rate of }(DR)_n$          &1/2 &1/2 &1/2 &1/2 &1/2 &1/2 &1/2 &1/2 &1/2 &1/2 & 1/2 &1/2\\\hline
$\mbox{ Rate of }TJC_n$          &1/2 &1/2 &1/2 &1/2 &1/2 &1/2 &1/2 &1/2 &1/2 &1/2 &1/2 &1/2\\\hline
$\mbox{ Rate of } L_n$          &2/3 &2/3 &5/8 &5/8 &3/5 &3/5 &7/12 &7/12 &4/7 &4/7 &9/16 &9/16 \\\hline\hline

\vspace*{0.2cm}
\end{tabular}
\end{center}
\end{table*}

\item As a byproduct of the above mentioned  construction, a general construction of  square Real Orthogonal Designs (RODs) is presented which includes as special cases well known constructions of Adams, Lax and Phillips \cite{ALP} and Geramita and Pullman \cite{GeP}.

\item Even though scaling only a subset of columns allowed us to decrease the delay, it is shown that such a scaling limits the rate of the design strictly to $\frac{1}{2}.$ In other words, the maximal rate of the scaled-CODs is $\frac{1}{2}$ when scaling is present in atleast one column.

\item Zero entries in a design increase the Peak-to-Average Power Ratio (PAPR) in the transmitted signal and it is preferred not to have any zero entries in the design. This problem has been addressed for square designs in \cite{DaR1,DaR2}. All the known maximal rate non-square designs have zero entries.  Our initial construction of rate-$\frac{1}{2}$ scaled CODs have zero entries in the design matrix which will lead to higher Peak-to-Average Power Ratio (PAPR) in contrast to the designs $TJC_n.$ However, we show that by post-multiplication of appropriate matrices, our construction leads to designs with no zero entries without change in the parameters of the design.

\end{itemize}
The remaining part of the paper is organized as follows: In Section \ref{sec2}, we present the main  result of the paper given by Theorem \ref{rate12cod}.  Before this, construction of a new set of maximal-rate square RODs is given in Subsection \ref{subsec2-1}. In Subsection \ref{subsec2-2}, construction of two new sets of rate-1 RODs from the maximal-rate square RODs of Subsection \ref{subsec2-1} is presented and in Subsection \ref{subsec2-3}, construction of the low-delay rate-1/2 scaled-CODs is achieved using rate-1 RODs of the previous subsection. In Subsection \ref{subsec2-4}, it is shown that the maximal rate for scaled-CODs is $\frac{1}{2}.$  For the special case of 9 transmit antennas, in Section \ref{sec3}, it is shown that our construction is of  minimal delay. In Section \ref{sec4}, we show that the codes discussed so far can be made to have no zero entries in by appropriate preprocessing without affecting the parameters of the design. Concluding remarks constitute Section \ref{sec5}. In Appendix \ref{appendixII}, it is shown that the well known constructions of square RODs by Adams-Lax-Phillips and Geramita-Pullman are special cases of our construction.
\section{A Construction of Rate-1/2 Scaled Complex Orthogonal Designs}
\label{sec2}
In this section, we construct a rate-1/2 scaled CODs for any number of transmit antennas with $50\%$ reduction in decoding delay compared to the rate-1/2 codes constructed by Tarokh et al \cite{TJC}. The construction of these codes is done in the following three steps:\\
STEP 1:  Construction of a new set of square RODs (Subsection \ref{subsec2-1}).\\
STEP 2:  Construction of two new sets of rate-1 RODs from the square RODs of STEP 1 (Subsection \ref{subsec2-2}). \\
STEP 3:  Construction of low-delay rate-1/2 scaled CODs using rate-1 RODs (Subsection \ref{subsec2-3}).

Before explaining these steps, we first build up some preliminary results needed to describe these steps. 

Let $\mathbb{F}_2=\{0,1\}$ be the finite field with two elements with addition and multiplication denoted by $b_1\oplus b_2$ and $b_1b_2$  for $b_1,b_2\in \mathbb{F}_2.$ We consider logical operations also on the elements of this field:  $b_1 +b_2$ and $\bar{b}_1$ represent respectively the logical disjunction (OR) of $b_1$ and $b_2$ and complement or negation of $b_1$, i.e., 
\begin{eqnarray}
\begin{array}{c}
\label{binaryop}
b_1+b_2=b_1\oplus b_2 \oplus b_1b_2,\\
\bar{b_1}=1\oplus b_1.
 \end{array}
\end{eqnarray}
For any finite subset  $B$ of the set of natural numbers $\mathbb{N}$,

let $a\in\mathbb{N}$ be the least integer such that $b < 2^a$ for any $b \in B.$ Often we identify each element of $B$ with an element of $\mathbb{F}_2^a$ using the following correspondence:
$b\in B \leftrightarrow (b_{a-1},\cdots,b_0)\in\mathbb{F}_2^a$ such that $b=\sum_{j=0}^{a-1} b_j2^j,b_j\in \mathbb{F}_2$.
The all zero vector and all one vector in $\mathbb{F}_2^a$ are denoted by $\mathbf{0}$ and $\mathbf{1}$ respectively. For $x\in B$, $\overline{x}, x^c$ and $\vert x\vert$ represent respectively the $2's$ complement of $x$ in $\mathbb{F}_2^a$, $1's$ complement of $x$ in $\mathbb{F}_2^a$ and Hamming weight of $x$. In other words,  $\overline{x}=2^a-x$ and $x^c=2^a-1-x.$ Let $x=(x_{a-1},\cdots,x_0)$ and $y=(y_{a-1},\cdots,y_0)$. Then, $x\oplus y, x\cdot y$ denote the component-wise modulo-2 addition and component-wise  multiplication (AND operation)  of $x$ and $y$ respectively i.e.,
$x\oplus y=(x_{a-1}\oplus y_{a-1},\cdots,x_0\oplus y_0)$ and $x\cdot y=(x_{a-1}y_{a-1},\cdots,x_0y_0)$. 

Let $Z_{2^a}=\{0,1,\cdots,2^a-1\}$.
For a set  $K\subset Z_{2^a}$, we define $\overline{K}=\{\overline{x}~\vert~ x\in K\}$, $K^c=\{x^c~\vert~ x\in K\}$,
$m \oplus K=\{m\oplus a~\vert~ a\in K\}$ for any $m\in Z_{2^a}$ and  $\vert K\vert$ denotes the number of elements in the set $K$.

For any two sets  $A,B$ with $B\subset A$, the set $A\setminus B$, consists of those elements of $A$, which are not in $B$. For two integers $i,j$, we use the notation $i\equiv j$, to indicate that $i-j=0 \mod 2.$

For any matrix of size $n_1\times n_2,$ the rows and columns of the matrix are labeled by the elements of $\{0,1,\cdots,n_1-1\}$ and $\{0,1,\cdots,n_2-1\}$ respectively. If $M$ is a $p\times n$ matrix in $k$ real variables $x_0,x_1, x_2,\cdots, x_{k-1}$, such that each non-zero entry of the matrix is $x_i$ or $-x_i$ for some $i\in\{0,1,\cdots,k-1\}$, obviously, it is not necessary that $M$ is a ROD. For example, 
$\begin{bmatrix}
    x_0   &x_1    \\
    x_1   & x_0  
\end{bmatrix}
$ is not a ROD. In the following (Lemma \ref{proper}), we derive a necessary and sufficient condition for the matrix $M$  to be a ROD.\\
A submatrix $M_2$ of size $2\times 2,$  constructed by choosing any two rows and any two columns of $M$ is called {\textit{proper}} if
\begin{itemize}
\item None of the entries of $M_2$ is zero and
\item It contains exactly two distinct variables.
\end{itemize}
\begin{example}
Consider the following matrix in three real variables $x_0,x_1$ and $x_2$
{\small
\begin{equation}
 \begin{bmatrix}
    x_0   &-x_1   &-x_2  & 0     \\
    x_1   & x_0   & 0    &-x_2     \\
    x_2   & 0     & x_0  & x_1     \\
     0    & x_2   &-x_1  & x_0
\end{bmatrix}.
\end{equation}}
The sub-matrix 
$\begin{bmatrix}
    x_1   &-x_2    \\
    x_2   & x_1   
  \end{bmatrix}
$ is {\textit{proper}} while
$
\begin{bmatrix}
    x_3   & 0 \\
    0     & x_3 
  \end{bmatrix}
$
is not.
\end{example}
%
%
If $M(i,j)\neq 0$, then we write $\vert M(i,j)\vert= k$ whenever $M(i,j)=x_k$ or $-x_k$ ( in case of CODs, $\vert M(i,j)\vert= k$ if $M(i,j)\in \{\pm x_k,\pm x_k^*\}$).

If $M$ is a ROD and if $M_2$ is a $2\times 2$ {\textit{proper}} sub-matrix of $M$, containing two variables, say $x_l$ and $x_m$, $l,m$ positive integers, then $M_2^\mathcal{T}M_2= (x_l^2+x_m^2)I_2$. In other words, $M_2$ is
a ROD by itself in two variables. The following lemma gives a characterization of RODs in term of proper $2\times 2$ matrices.

\begin{lemma}
\label{proper}
Let $M$ be a $p\times n$ matrix in $k$ real variables $x_0,x_1, x_2,\cdots, x_{k-1}$, such that each non-zero entry of the matrix is $x_i$ or $-x_i$ for some $i\in\{0,1,\cdots,k-1\}$. Then the following two statements are equivalent:\\
 1) $M$ is a ROD.\\
 2) (i) Each variable appears exactly once along each column of $M$ and atmost once along each row of $M$,\\
    (ii) if for some $i,j,j^\prime$, $M(i,j)\neq 0$ and $M(i,j^\prime)\neq 0$, then
         there exists $i^\prime$ such that $\vert M(i,j)\vert=\vert M(i^\prime,j^\prime)\vert$ and $\vert M(i,j^\prime)\vert=\vert M(i^\prime,j)\vert$,\\
   (iii) any proper $2\times 2$ sub-matrix of $M$ is a ROD.
\end{lemma}
\subsection{STEP 1: Construction of a new class of square RODs}
\label{subsec2-1}
Square RODs have been constructed by several authors, for example, Adams et al \cite{ALP} and Geramita et al \cite{GeP}. All these designs are constructed recursively and the basic blocks of these designs are the RODs of order $1,2,4$ and $8$. It is known that these designs are obtained by left (or right) regular representations of the field of real numbers, the field of complex numbers, the Quaternion algebra and the Octonion algebra respectively. In this subsection, we take a different approach towards the construction of RODs and that lead to a new class of RODs constructable recursively of which the constructions of \cite{ALP} and \cite{GeP} are special cases.
  
If $B_t$ is a square real design of size $[t,t,k]$ in $k$ real variables $x_0,\cdots,x_{k-1}$, then whenever $B_t(i,j)\neq 0$, we write
\begin{eqnarray}
\begin{array}{c}
\label{definesf1}
B_t(i,j)={\mu}_t(i,j)x_{\lambda_t(i,j)},\\
 \mbox{ for some } \hspace{1.00cm} {\mu}_t(i,j)\in\{1,-1\},\\
\mbox{  and } \hspace{1.00cm}  \lambda_t(i,j)\in\{0,1,\cdots,k-1\}\\\
 \mbox{ for } 0\leq i,j\leq t-1.
 \end{array}
\end{eqnarray}
$B_t$ is uniquely determined by $\mu_t$ and $\lambda_t$.\\

The approach we take is identifying a pair of functions $\mu_t$ and $\lambda_t$ that defines a square ROD. Towards that end, for any  $t,$ identifying $Z_{2^a}$ with $\mathbb{F}_2^{a},$ we have $S\subset Z_{2^a}$ identified with a subset of $\mathbb{F}_2^{a}$. 
We define two maps which are used for the construction of our square RODs as follows. \\
Let 
\begin{equation}
\label{gammamap}
\gamma_t: Z_{\rho(t)}\rightarrow Z_t 
\end{equation}
be an injective map defined on $Z_{\rho(t)}$ with the image denoted by
$\hat{Z}_{\rho(t)}=\gamma_t(Z_{\rho(t)})$ 
and  
\begin{equation}
\label{psimap}
\psi_t: \hat{Z}_{\rho(t)}\rightarrow Z_t 
\end{equation}
be another injective map defined on $\hat{Z}_{\rho(t)}.$ 

In the following theorem, we define the two maps $\mu_t$ and $\lambda_t$ given in \eqref{definesf1} which define the matrix $B_t,$ in terms of the two maps \eqref{gammamap} and \eqref{psimap} and identify the conditions for the resulting $B_t$ to be a square ROD. 
\begin{theorem}
\label{ratesquarerod}
Let $t=2^a$ and $B_t$ be a real design with $B_t(i,j)$ be non-zero if and only if $i\oplus j\in \hat{Z}_{\rho(t)}$. When $B_t(i,j)\neq 0$, let
\begin{eqnarray}
\label{definesf}
\mu_t(i,j)&=&(-1)^{\vert i\cdot\psi_t(i\oplus j)\vert},\\
\lambda_t(i,j)&=&\gamma_t^{-1}(i\oplus j).
\end{eqnarray}
If 
\begin{equation}
\label{oddcondition}
\vert (\psi_t(x)\oplus \psi_t(y))\cdot(x\oplus y)\vert
\end{equation}
is odd, for all $x,y\in\hat{Z}_{\rho(t)}, x\neq y,$ then, $B_t$ is a square ROD of size $[t,t,\rho(t)]$.
\end{theorem}
\begin{proof}
We use the Lemma~\ref{proper} to prove that $B_t$ is a ROD.
First, for a fixed $j\in Z_t,$ define
\[A=\{i\oplus j~\vert i\in Z_t, i\oplus j\in \hat{Z}_{\rho(t)}\}.\]
It is clear that $A=\hat{Z}_{\rho(t)}$.
Moreover, as $\gamma_t$ is injective, we have $\gamma_t^{-1}(i\oplus j)\neq \gamma_t^{-1}(i^\prime\oplus j)$ whenever $i\neq i^\prime$.
Therefore, each column of the matrix $B_t$ contains all the variables $x_0,x_1,\cdots x_{\rho(t)-1}$ and these variables appear exactly once.
Similarly, it follows that the variables appear atmost once in any row of $B_t$.

Secondly, assume that $B_t(i,j)\neq 0$ and $B_t(i,j^\prime)\neq 0$, then we show that there exists $i^\prime$ such that
\begin{eqnarray*}
\vert B_t(i,j)\vert=\vert B_t(i^\prime,j^\prime)\vert,\\
\vert B_t(i,j^\prime)\vert = \vert B_t(i^\prime,j)\vert.
\end{eqnarray*}

Let $i^\prime=i\oplus j\oplus j^\prime$. We have
\begin{eqnarray*}
\vert B_t(i,j)\vert=\gamma_t^{-1}(i\oplus j),\\
\vert B_t(i^\prime,j^\prime)\vert=\gamma_t^{-1}(i^\prime\oplus j^\prime)=\gamma_t^{-1}(i\oplus j).
\end{eqnarray*}
Therefore, $\vert B_t(i,j)\vert=\vert B_t(i^\prime,j^\prime)\vert$.
Similarly, $\vert B_t(i,j^\prime)\vert=\vert B_t(i^\prime,j)\vert$.
Thirdly, we show that any proper $2\times 2$ sub-matrix of $B_t$ is a ROD. It is enough to prove that
$\mu_t(i,j)\cdot \mu_t(i,j^\prime)\cdot \mu_t(i^\prime,j)\cdot \mu_t(i^\prime,j^\prime)=-1$,
or equivalently,
\begin{equation}
\label{inter}
\vert i\cdot\psi_t(i\oplus j) \vert +\vert i\cdot\psi_t(i\oplus j^\prime) \vert
+\vert i^\prime\cdot\psi_t(i^\prime\oplus j) \vert+\vert i^\prime\cdot\psi_t(i^\prime\oplus j^\prime)\vert
\end{equation}
is an odd number.
But $i^\prime=i\oplus j\oplus j^\prime$ and $\vert x\oplus y \vert \equiv\vert x\vert +\vert y \vert$
where $k\equiv l$ if $k-l$ is a multiple of $2$.
We can write \eqref{inter} as
\begin{eqnarray*}
\begin{array}{r}
\vert (i\oplus i^\prime)\cdot(\psi_t(i\oplus j) \oplus \psi_t(i^\prime\oplus j)) \vert\\
=\vert \bigl((i\oplus j)\oplus(i^\prime\oplus j))\cdot(\psi_t(i\oplus j) \oplus \psi_t(i^\prime\oplus j)\bigr) \vert.
\end{array}
\end{eqnarray*}
which is an odd number as both $i\oplus j$ and $i^\prime\oplus j$ are the elements of $\hat{Z}_{\rho(t)}$.
\end{proof}

Thus, it is enough to construct $\gamma_t$ and $\psi_t$ satisfying the property stated in Theorem \ref{ratesquarerod}, in order to construct a square ROD. This we achieve by making use of another set of two maps $\phi_1$ and $\phi_2$ as follows.

On $Z_8=\{0,1,\cdots,7\}$ we define a map  $\phi_1: Z_8\mapsto Z_8$ given by
 \begin{equation}  
\label{phi_07}
\phi_1=\left(\begin{array}{cccccccccccc}
    0   & 1 &2 &3&4&5&6&7   \\
    0   & 1 &2 &3&4&7&5&6
    \end{array}\right).
\end{equation}
For $a\in \{0,1,2,3\}$, we have $Z_{2^a}\subseteq Z_8$.
Define
\begin{eqnarray}
\label{psi1248}
\hat{\psi}_{2^a}: Z_{2^a}\mapsto Z_{2^a}\nonumber\\
    x\mapsto \overline{\phi_1(x)}
\end{eqnarray}
where the $2$'s complement of an element $x$ of $\mathbb{F}_2^a$ is performed in $\mathbb{F}_2^a.$ Note that the map $\hat{\psi}_{2^a}$ is well defined even though it is defined in terms of $\phi_1$ which is defined on $Z_8.$ 
\begin{lemma}
\label{propertyV8}
Let $x,y\in Z_{2^a},a\in\{0,1,2,3\}, x\neq y$. Identify $Z_{2^a}$ with $\mathbb{F}_2^a$. Then $\vert (\hat{\psi}_{2^a}(x)\oplus \hat{\psi}_{2^a}(y))\cdot(x\oplus y)\vert$  is an odd integer. 
\end{lemma}
\begin{proof}
We prove it only for $a=3$.
For all other values of $a$, one can prove it similarly.
Write $\hat{x}=\hat{\psi}_8(x)$.
Let the radix-2 representation of $x\in Z_8$ be $(x_2,x_1,x_0)$ and  that of $\hat{x}$ be $(\hat{x}_2,\hat{x}_1,\hat{x}_0)$ where each $x_i$ or $\hat{x}_i$ takes value from the set $\mathbb{F}_2=\{0,1\}$.
The following table describes the map $x\rightarrow \hat{x}$ for $x=0,\cdots,7$.
\begin{equation*}
\label{truthtable}
\begin{tabular}{|c||rrr||rrr||}\hline
$x$ & $x_2$&$x_1$&$x_0$&$\hat{x}_2$& $\hat{x}_1$& $\hat{x}_0$ \\ \hline 
0 &0 &0 &0  &0 &0 &0 \\ \hline
1 &0 &0 &1  &1 &1 &1 \\ \hline
2 &0 &1 &0  &1 &1 &0 \\ \hline
3 &0 &1 &1  &1 &0 &1 \\ \hline
4 &1 &0 &0  &1 &0 &0 \\ \hline
5 &1 &0 &1  &0 &0 &1 \\ \hline
6 &1 &1 &0  &0 &1 &1 \\ \hline
7 &1 &1 &1  &0 &1 &0 \\ \hline
\end{tabular}
\end{equation*}

Using this table, and \eqref{binaryop} we express $\hat{x}_i$ in term of $x_2,x_1$ and $x_0$ for $i=0,1$ and $2$ as follows.
\begin{eqnarray*}
\hat{x}_0&=&\bar{x}_2 x_0+x_2 (x_1\oplus x_0)=x_0\oplus x_1 x_2,\\
\hat{x}_1&=&x_2 x_1+\bar{x}_2 (x_1\oplus x_0)=x_0\oplus x_1 \oplus x_2 x_0,\\
\hat{x}_2&=&\bar{x}_2 x_1+\bar{x}_1(x_2\oplus x_0)=x_0\oplus x_1\oplus x_2 \oplus x_0 x_1.
\end{eqnarray*}

Hence
 \[\hat{x}_i=\sum_{j=0}^{i} x_j \oplus \prod_{\substack{0\leq j\leq 2\\j\neq i}}x_j,~~ i=0,1 \mbox{ and } 2.\]
Let $x,y\in Z_8, x\neq y$.
Now $\vert (\hat{x}\oplus \hat{y})\cdot(x \oplus y)\vert$ is odd
if and only if
\[\sum_{i=0}^{2}(x_i\oplus y_i) (\hat{x}_i\oplus\hat{y}_i)=1.\]
But $\sum_{i=0}^{2}(x_i\oplus y_i) (\hat{x}_i\oplus\hat{y}_i)=1\oplus\prod_{i=0}^2 (1\oplus x_i \oplus y_i)$.\\
Now $1\oplus\prod_{i=0}^2 (1\oplus x_i \oplus y_i)$ is equal to $0$ if and only if \\ $\prod_{j=0}^2 (1\oplus x_i \oplus y_i)=1$
i.e., $(x_i \oplus y_i)=0$ for all $i$, which implies that $x=y$.\\
As $x\neq y$, we have $\sum_{i=0}^{2}(x_i\oplus y_i) (\hat{x}_i\oplus\hat{y}_i)=1.$
\end{proof}

The second map $\phi_2$ is defined on the set $F$ given by 
\begin{eqnarray}
\label{defineef}
F=\Big\{x\in\mathbb{F}_2^4 \Big\vert~~ \vert x\vert =1 \mbox { or } 3 \Big\}=\{1,2,4,7,8,11,13,14 \}
\end{eqnarray}
as injective map 
$\phi_2: F\mapsto Z_{16}$ given by 
\begin{equation}  
\phi_2=\left(\begin{array}{cccccccccccc}
     1 &2 &4&7&8&11&13&14   \\
     1 &2 &4&6&8&15&10&12
    \end{array}\right).
\end{equation}
\begin{lemma}
\label{propertyphi}
Let $F$ be the set given in \eqref{defineef} and $x,y\in F, x\neq y$. Then \\
(i) $\vert \overline{\phi_2(x)}\cdot x\vert$  is odd for all $x\neq 0$.\\
(ii) $\vert \overline{\phi_2(x)}\cdot y\vert +
\vert \overline{\phi_2(y)}\cdot x\vert$ is  odd for all  $x\neq y,
x\neq 0,y\neq 0$.\\
\end{lemma}
\begin{proof}
There are only finitely many possibilities for $x$ and $y$ and it can be checked that both the statements (i) and (ii) hold for all possible cases.
\end{proof}

Now, we define the maps $\gamma_t$ and $\psi_t$ in terms of the maps $\phi_1$ and $\phi_2$ as follows:
The map $\gamma_t$  defined over $Z_{\rho(t)}$ is given by
\begin{eqnarray}
\label{mapgamma}
 \gamma_t(i)=
    \begin{cases}
i   & \text{ if  $0\leq i\leq 7$ } \\
2^{4l-1}\cdot\hat{\gamma}(m)&\text{ if $i\geq 8,i=8l+m$, $0\leq m\leq 7$ }
  \end{cases}
  \end{eqnarray}
where 
\begin{equation}  
\hat{\gamma}=\left(\begin{array}{cccccccc}
 0& 1 &2 &3&4&5&6&7 \\
 1& 2 &4 &7&8&11&13&14
\end{array}\right).
\end{equation}
For an element $x\in \hat{Z}_{\rho(t)}$, either $x\in Z_8$ or $x=2^{4y-1}z$ for some $y\in\mathbb{N}\setminus\{0\}$ and $z\in F$.
Let $\phi$ be the map defined on the set $\hat{Z}_{\rho(t)}$ given by
\begin{eqnarray}
\phi(x)&=&
    \begin{cases}
\phi_1(x)  & \text{ if  $x\in Z_8$} \\
2^{4y-1}\cdot\phi_2(z)&\text{ if $x=2^{4y-1}z,~z\in F$}.
  \end{cases}
  \end{eqnarray}
The map $\psi_t$  is defined by 
\begin{equation}
\label{mappsi}
\psi_t(x)=\overline{\phi(x)} \mbox{ in } \mathbb{F}_2^a~ \forall x\in \hat{Z}_{\rho(t)}.
\end{equation}

The following theorem shows that the maps $\gamma_t$ and $\psi_t$ defined by \eqref{mapgamma} and \eqref{mappsi} satisfy the conditions of Theorem \ref{ratesquarerod} and hence define a ROD.  
\begin{theorem}
\label{orthog}
Identify  $\hat{Z}_{\rho(t)}$  with a subset of $\mathbb{F}_2^{a}$, $t=2^a$. Then
$ \vert (\psi_t(x)\oplus \psi_t(y))\cdot(x\oplus y)\vert$  is odd for all $x,y\in\hat{Z}_{\rho(t)}, x\neq y,$ and hence from Theorem \ref{ratesquarerod} the matrix $B_t$ defined by $\gamma_t$ and $\psi_t$ by \eqref{mapgamma} and \eqref{mappsi} is a square ROD.  
\end{theorem}
\begin{proof}
For $t=1,2,4$ and $8$, the statement holds by Lemma \ref{propertyV8}. Hence we assume that
$t\geq 16$.
As $\psi_t(0)=0$, it is enough to prove that \\
(i) $\vert \psi_t(y)\cdot y\vert$ is odd for all $y\neq 0$.\\
(ii)$\vert\psi_t(x)\cdot y\vert +\vert\psi_t(y)\cdot x\vert$ is odd for all  $x\neq y,\\x\neq 0,y\neq 0$.\\
To prove (i),
let $z =\psi_t(y)\cdot y$. If $y\in E$, we have $\vert\psi_t(y)\cdot y\vert=\vert\psi_8(y)\cdot y\vert$ which is an odd number by Lemma \ref{propertyV8}.\\
On the other hand, if $y=2^{4l-1}m$, $l\geq 0,m\in F$, then
$\vert z \vert=\vert\overline{2^{4l-1}\phi_2(m)}\cdot 2^{4l-1}m\vert$ where the $2's$ complement of an element is performed in $\mathbb{F}_2^a$.
We have $\vert z \vert=\vert\overline{\phi_2(m)}\cdot m\vert$ where the $2's$ complement of $\phi_2(m)$ is performed in $\mathbb{F}_2^4$. Hence $\vert z \vert$ is odd by Lemma \ref{propertyphi}.

In order to prove the part (ii),
we have following three cases: \\
(i) $1\leq x\leq 7 ~\&~ 1\leq y\leq 7$, \\
(ii) $1\leq y\leq 7~\&~ x=2^{4\alpha-1}\beta$ for some $\beta\in F$, $\alpha\geq 1$,\\
(iii) $x=2^{4\hat{\alpha}-1}\hat{\beta}~\&~ y=2^{4\alpha-1}\beta$ for some $\beta,\hat{\beta}\in F,\alpha,\hat{\alpha}\geq 1$.\\
In all the three cases, we have $x\neq y$.
By Lemma \ref{propertyV8}, (i) is true.  \\ 
For the second case, let $z=\psi_t(x)\cdot y \oplus \psi_t(y)\cdot x$.
We have $z=(\overline{2^{4\alpha-1}\phi_2(\beta)}\cdot y) \oplus ((2^{4\alpha-1}\beta)\cdot\overline{\phi_1(y)})$.

As $\overline{2^{4\alpha-1}\phi_2(\beta)}\cdot y=\mathbf{0}$ (the all zero vector in $\mathbb{F}_2^{a}$) for $\alpha\geq 1$, we have
$z=(2^{4\alpha-1}\beta)\cdot\overline{\phi_1(y)}$.
But $\vert\beta\vert$ is odd  for all $\beta\in F$, hence $\vert z \vert$ is an odd number.

For (iii), let $z=\psi_t(x)\cdot y \oplus \psi_t(y)\cdot x$.
We have
\[z=\overline{2^{4\alpha-1}\phi_2(\beta)}\cdot 2^{4\hat{\alpha}-1}\hat{\beta}
\oplus 2^{4\alpha-1}\beta\cdot \overline{2^{4\hat{\alpha}-1}\phi_2(\hat{\beta})}.
\]
If $\hat{\alpha}>\alpha$, we have $2^{4\alpha-1}\beta\cdot \overline{2^{4\hat{\alpha}-1}\phi_2(\hat{\beta})}=\mathbf{0}$ and
$\overline{2^{4\alpha-1}\phi_2(\beta)}\cdot 2^{4\hat{\alpha}-1}\hat{\beta}=\hat{\beta}$. Thus $\vert z\vert$ is an odd number by Lemma \ref{propertyphi}.
If $\alpha=\hat{\alpha}$, it follows that
\begin{equation*}
\vert z\vert = \vert \overline{\phi_2(\beta)}\cdot \hat{\beta}\vert +
\vert \beta\cdot \overline{\phi_2(\hat{\beta})}\vert
\end{equation*}
which is an odd number by Lemma \ref{propertyphi}. 

From Theorem \ref{ratesquarerod} it follows that the matrix $B_t$ defined by $\gamma_t$ and $\psi_t$ by \eqref{mapgamma} and \eqref{mappsi} is a square ROD. 
\end{proof}

The square RODs of Theorem \ref{orthog} will be denoted by $R_t$ throughout. The ROD $R_{16}$ of size $[16,16,9]$ is given by \eqref{R16} at the top of the next page. As another example the ROD $R_{32}$ of size $[32,32,10]$ is given by \eqref{R32} in the following page. In Appendix \ref{appendixI}, it is shown that the RODs $R_t$ can be constructed recursively.
\begin{figure*}
\begin{eqnarray}
\label{R16}
R_{16}=
\left[\begin{array}{rrrrrrrrrrrrrrrr}
x_0&x_1&  x_2& x_3& x_4& x_5&  x_6 & x_7 & x_8 &  0 &  0 &  0 &  0 &  0 &  0 &  0\\
\vspace{-0.1cm}
 -x_1 & x_0& -x_3 & x_2& -x_5 & x_4 & x_7 & -x_6 &  0 & x_8 &  0 &  0 &  0 &  0 &  0 &  0 \\
\vspace{-0.1cm}
 -x_2 & x_3 & x_0 & -x_1 & -x_6 & -x_7 & x_4 & x_5 &  0 &  0 & x_8 &  0 &  0 &  0 &  0 &  0 \\
\vspace{-0.1cm}
 -x_3 & -x_2 & x_1 & x_0 & -x_7 & x_6 & -x_5 & x_4 &  0 &  0 &  0 & x_8 &  0 &  0 &  0 &  0 \\
\vspace{-0.1cm}
 -x_4 & x_5 & x_6 & x_7 & x_0 & -x_1 & -x_2 & -x_3 &  0 &  0 &  0 &  0 & x_8 &  0 &  0 &  0 \\
\vspace{-0.1cm}
 -x_5 & -x_4 & x_7 & -x_6 & x_1 & x_0 & x_3 & -x_2 &  0 &  0 &  0 &  0 &  0 & x_8 &  0 &  0 \\
\vspace{-0.1cm}
 -x_6 & -x_7 & -x_4 & x_5 & x_2 & -x_3 & x_0 & x_1 &  0 &  0 &  0 &  0 &  0 &  0 & x_8 &  0 \\
\vspace{-0.1cm}
 -x_7 & x_6 & -x_5 & -x_4 & x_3 & x_2 & -x_1 & x_0 &  0 &  0 &  0 &  0 &  0 &  0 &  0 & x_8 \\
\vspace{-0.1cm}
 -x_8 &  0 &  0 &  0 &  0 &  0 &  0 &  0 & x_0 & -x_1 & -x_2 & -x_3 & -x_4 & -x_5 & -x_6 & -x_7 \\
\vspace{-0.1cm}
   0 & -x_8 &  0 &  0 &  0 &  0 &  0 &  0 & x_1 & x_0 & x_3 & -x_2 & x_5 & -x_4 & -x_7 & x_6 \\
\vspace{-0.1cm}
   0 &  0 & -x_8 &  0 &  0 &  0 &  0 &  0 & x_2 & -x_3 & x_0 & x_1 & x_6 & x_7 & -x_4 & -x_5 \\
\vspace{-0.1cm}
   0 &  0 &  0 & -x_8 &  0 &  0 &  0 &  0 & x_3 & x_2 & -x_1 & x_0 & x_7 & -x_6 & x_5 & -x_4 \\
\vspace{-0.1cm}
   0 &  0 &  0 &  0 & -x_8 &  0 &  0 &  0 & x_4 & -x_5 & -x_6 & -x_7 & x_0 & x_1 & x_2 & x_3 \\
\vspace{-0.1cm}
   0 &  0 &  0 &  0 &  0 & -x_8 &  0 &  0 & x_5 & x_4 & -x_7 & x_6 & -x_1 & x_0 & -x_3 & x_2 \\
\vspace{-0.1cm}
   0 &  0 &  0 &  0 &  0 &  0 & -x_8 &  0 & x_6 & x_7 & x_4 & -x_5 & -x_2 & x_3 & x_0 & -x_1 \\
\vspace{-0.1cm}
   0 &  0 &  0 &  0 &  0 &  0 &  0 & -x_8 & x_7 & -x_6 & x_5 & x_4 & -x_3 & -x_2 & x_1 & x_0 
\end{array}\right]
\end{eqnarray}
\hrule
\end{figure*}
\begin{figure*}
\tiny{
\begin{eqnarray}
\label{R32}
R_{32}=\left[\begin{array}{ r @{\hspace{.2pt}} r @{\hspace{.2pt}} r @{\hspace{.2pt}} r @{\hspace{.2pt}}r @{\hspace{.2pt}} r @{\hspace{.2pt}} r @{\hspace{.2pt}} r @{\hspace{.2pt}}
r @{\hspace{.2pt}} r @{\hspace{.2pt}} r @{\hspace{.2pt}} r @{\hspace{.2pt}}r @{\hspace{.2pt}} r @{\hspace{.2pt}} r @{\hspace{.2pt}} r @{\hspace{.2pt}} r @{\hspace{.2pt}} r @{\hspace{.2pt}} r @{\hspace{.2pt}} r @{\hspace{.2pt}}r @{\hspace{.2pt}} r @{\hspace{.2pt}} r @{\hspace{.2pt}} r @{\hspace{.2pt}}r @{\hspace{.2pt}} r @{\hspace{.2pt}} r @{\hspace{.2pt}} r @{\hspace{.2pt}}r @{\hspace{.2pt}} r @{\hspace{.2pt}} r @{\hspace{.2pt}} r @{\hspace{.2pt}}}
  x_0 & x_1 & x_2 & x_3 & x_4 & x_5 & x_6 & x_7 & x_8 &  0 &  0 &  0 &  0 &  0 &  0 &  0 & x_9 &  0 &  0 &  0 &  0 &  0 &  0 &  0 &  0 &  0 &  0 &  0 &  0 &  0 &  0 &  0 \\
 -x_1 & x_0 & -x_3 & x_2 & -x_5 & x_4 & x_7 & -x_6 &  0 & x_8 &  0 &  0 &  0 &  0 &  0 &  0 & 0 & x_9 &  0 &  0 &  0 &  0 &  0 &  0 &  0 &  0 &  0 &  0 &  0 &  0 &  0 &  0 \\
 -x_2 & x_3 & x_0 & -x_1 & -x_6 & -x_7 & x_4 & x_5 &  0 &  0 & x_8 &  0 &  0 &  0 &  0 &  0 & 0 &  0 & x_9 &  0 &  0 &  0 &  0 &  0 &  0 &  0 &  0 &  0 &  0 &  0 &  0 &  0 \\
 -x_3 & -x_2 & x_1 & x_0 & -x_7 & x_6 & -x_5 & x_4 &  0 &  0 &  0 & x_8 &  0 &  0 &  0 &  0 & 0 &  0 &  0 & x_9 &  0 &  0 &  0 &  0 &  0 &  0 &  0 &  0 &  0 &  0 &  0 &  0 \\
 -x_4 & x_5 & x_6 & x_7 & x_0 & -x_1 & -x_2 & -x_3 &  0 &  0 &  0 &  0 & x_8 &  0 &  0 &  0 & 0 &  0 &  0 &  0 & x_9 &  0 &  0 &  0 &  0 &  0 &  0 &  0 &  0 &  0 &  0 &  0 \\
 -x_5 & -x_4 & x_7 & -x_6 & x_1 & x_0 & x_3 & -x_2 &  0 &  0 &  0 &  0 &  0 & x_8 &  0 &  0 & 0 &  0 &  0 &  0 &  0 & x_9 &  0 &  0 &  0 &  0 &  0 &  0 &  0 &  0 &  0 &  0 \\
 -x_6 & -x_7 & -x_4 & x_5 & x_2 & -x_3 & x_0 & x_1 &  0 &  0 &  0 &  0 &  0 &  0 & x_8 &  0 & 0 &  0 &  0 &  0 &  0 &  0 & x_9 &  0 &  0 &  0 &  0 &  0 &  0 &  0 &  0 &  0 \\
 -x_7 & x_6 & -x_5 & -x_4 & x_3 & x_2 & -x_1 & x_0 &  0 &  0 &  0 &  0 &  0 &  0 &  0 & x_8 & 0 &  0 &  0 &  0 &  0 &  0 &  0 & x_9 &  0 &  0 &  0 &  0 &  0 &  0 &  0 &  0 \\
 -x_8 &  0 &  0 &  0 &  0 &  0 &  0 &  0 & x_0 & -x_1 & -x_2 & -x_3 & -x_4 & -x_5 & -x_6 & -x_7 &  0 &  0 &  0 &  0 &  0 &  0 &  0 &  0 & x_9 &  0 &  0 &  0 &  0 &  0 &  0 &  0 \\
   0 & -x_8 &  0 &  0 &  0 &  0 &  0 &  0 & x_1 & x_0 & x_3 & -x_2 & x_5 & -x_4 & -x_7 & x_6 &  0 &  0 &  0 &  0 &  0 &  0 &  0 &  0 &  0 & x_9 &  0 &  0 &  0 &  0 &  0 &  0 \\
   0 &  0 & -x_8 &  0 &  0 &  0 &  0 &  0 & x_2 & -x_3 & x_0 & x_1 & x_6 & x_7 & -x_4 & -x_5 &  0 &  0 &  0 &  0 &  0 &  0 &  0 &  0 &  0 &  0 & x_9 &  0 &  0 &  0 &  0 &  0 \\
   0 &  0 &  0 & -x_8 &  0 &  0 &  0 &  0 & x_3 & x_2 & -x_1 & x_0 & x_7 & -x_6 & x_5 & -x_4 &  0 &  0 &  0 &  0 &  0 &  0 &  0 &  0 &  0 &  0 &  0 & x_9 &  0 &  0 &  0 &  0 \\
   0 &  0 &  0 &  0 & -x_8 &  0 &  0 &  0 & x_4 & -x_5 & -x_6 & -x_7 & x_0 & x_1 & x_2 & x_3 &  0 &  0 &  0 &  0 &  0 &  0 &  0 &  0 &  0 &  0 &  0 &  0 & x_9 &  0 &  0 &  0 \\
   0 &  0 &  0 &  0 &  0 & -x_8 &  0 &  0 & x_5 & x_4 & -x_7 & x_6 & -x_1 & x_0 & -x_3 & x_2 &  0 &  0 &  0 &  0 &  0 &  0 &  0 &  0 &  0 &  0 &  0 &  0 &  0 & x_9 &  0 &  0 \\
   0 &  0 &  0 &  0 &  0 &  0 & -x_8 &  0 & x_6 & x_7 & x_4 & -x_5 & -x_2 & x_3 & x_0 & -x_1 &  0 &  0 &  0 &  0 &  0 &  0 &  0 &  0 &  0 &  0 &  0 &  0 &  0 &  0 & x_9 &  0 \\
   0 &  0 &  0 &  0 &  0 &  0 &  0 & -x_8 & x_7 & -x_6 & x_5 & x_4 & -x_3 & -x_2 & x_1 & x_0 &  0 &  0 &  0 &  0 &  0 &  0 &  0 &  0 &  0 &  0 &  0 &  0 &  0 &  0 &  0 & x_9 \\
 -x_9 &  0 &  0 &  0 &  0 &  0 &  0 &  0 &  0 &  0 &  0 &  0 &  0 &  0 &  0 &  0 & x_0 & -x_1 & -x_2 & -x_3 & -x_4 & -x_5 & -x_6 & -x_7 & -x_8 &  0 &  0 &  0 &  0 &  0 &  0 &  0 \\
   0 & -x_9 &  0 &  0 &  0 &  0 &  0 &  0 &  0 &  0 &  0 &  0 &  0 &  0 &  0 &  0 & x_1 & x_0 & x_3 & -x_2 & x_5 & -x_4 & -x_7 & x_6 &  0 & -x_8 &  0 &  0 &  0 &  0 &  0 &  0 \\
   0 &  0 & -x_9 &  0 &  0 &  0 &  0 &  0 &  0 &  0 &  0 &  0 &  0 &  0 &  0 &  0 & x_2 & -x_3 & x_0 & x_1 & x_6 & x_7 & -x_4 & -x_5 &  0 &  0 & -x_8 &  0 &  0 &  0 &  0 &  0 \\
   0 &  0 &  0 & -x_9 &  0 &  0 &  0 &  0 &  0 &  0 &  0 &  0 &  0 &  0 &  0 &  0 & x_3 & x_2 & -x_1 & x_0 & x_7 & -x_6 & x_5 & -x_4 &  0 &  0 &  0 & -x_8 &  0 &  0 &  0 &  0 \\
   0 &  0 &  0 &  0 & -x_9 &  0 &  0 &  0 &  0 &  0 &  0 &  0 &  0 &  0 &  0 &  0 & x_4 & -x_5 & -x_6 & -x_7 & x_0 & x_1 & x_2 & x_3 &  0 &  0 &  0 &  0 & -x_8 &  0 &  0 &  0 \\
   0 &  0 &  0 &  0 &  0 & -x_9 &  0 &  0 &  0 &  0 &  0 &  0 &  0 &  0 &  0 &  0 & x_5 & x_4 & -x_7 & x_6 & -x_1 & x_0 & -x_3 & x_2 &  0 &  0 &  0 &  0 &  0 & -x_8 &  0 &  0 \\
   0 &  0 &  0 &  0 &  0 &  0 & -x_9 &  0 &  0 &  0 &  0 &  0 &  0 &  0 &  0 &  0 & x_6 & x_7 & x_4 & -x_5 & -x_2 & x_3 & x_0 & -x_1 &  0 &  0 &  0 &  0 &  0 &  0 & -x_8 &  0 \\
   0 &  0 &  0 &  0 &  0 &  0 &  0 & -x_9 &  0 &  0 &  0 &  0 &  0 &  0 &  0 &  0 & x_7 & -x_6 & x_5 & x_4 & -x_3 & -x_2 & x_1 & x_0 &  0 &  0 &  0 &  0 &  0 &  0 &  0 & -x_8 \\
   0 &  0 &  0 &  0 &  0 &  0 &  0 &  0 & -x_9 &  0 &  0 &  0 &  0 &  0 &  0 &  0 & x_8 &  0 &  0 &  0 &  0 &  0 &  0 &  0 & x_0 & x_1 & x_2 & x_3 & x_4 & x_5 & x_6 & x_7 \\
   0 &  0 &  0 &  0 &  0 &  0 &  0 &  0 &  0 & -x_9 &  0 &  0 &  0 &  0 &  0 &  0 &  0 & x_8 &  0 &  0 &  0 &  0 &  0 &  0 & -x_1 & x_0 & -x_3 & x_2 & -x_5 & x_4 & x_7 & -x_6 \\
   0 &  0 &  0 &  0 &  0 &  0 &  0 &  0 &  0 &  0 & -x_9 &  0 &  0 &  0 &  0 &  0 &  0 &  0 & x_8 &  0 &  0 &  0 &  0 &  0 & -x_2 & x_3 & x_0 & -x_1 & -x_6 & -x_7 & x_4 & x_5 \\
   0 &  0 &  0 &  0 &  0 &  0 &  0 &  0 &  0 &  0 &  0 & -x_9 &  0 &  0 &  0 &  0 &  0 &  0 & 0 & x_8 &  0 &  0 &  0 &  0 & -x_3 & -x_2 & x_1 & x_0 & -x_7 & x_6 & -x_5 & x_4 \\
   0 &  0 &  0 &  0 &  0 &  0 &  0 &  0 &  0 &  0 &  0 &  0 & -x_9 &  0 &  0 &  0 &  0 &  0 & 0 &  0 & x_8 &  0 &  0 &  0 & -x_4 & x_5 & x_6 & x_7 & x_0 & -x_1 & -x_2 & -x_3 \\
   0 &  0 &  0 &  0 &  0 &  0 &  0 &  0 &  0 &  0 &  0 &  0 &  0 & -x_9 &  0 &  0 &  0 &  0 & 0 &  0 &  0 & x_8 &  0 &  0 & -x_5 & -x_4 & x_7 & -x_6 & x_1 & x_0 & x_3 & -x_2 \\
   0 &  0 &  0 &  0 &  0 &  0 &  0 &  0 &  0 &  0 &  0 &  0 &  0 &  0 & -x_9 &  0 &  0 &  0 & 0 &  0 &  0 &  0 & x_8 &  0 & -x_6 & -x_7 & -x_4 & x_5 & x_2 & -x_3 & x_0 & x_1 \\
   0 &  0 &  0 &  0 &  0 &  0 &  0 &  0 &  0 &  0 &  0 &  0 &  0 &  0 &  0 & -x_9 &  0 &  0 & 0 &  0 &  0 &  0 &  0 & x_8 & -x_7 & x_6 & -x_5 & -x_4 & x_3 & x_2 & -x_1 & x_0
\end{array}\right]
\end{eqnarray}
}
\hrule
\end{figure*}

One can define the functions $\gamma_t$ and $\psi_t$ different from the one given above and can have a square ROD different from $R_t$. In Appendix \ref{appendixII}, we provide three different pairs of such functions  and these are shown to give the well known  Adams-Lax-Phillips' construction from Octonions and Quaternions and Geramita and Pullman's construction of square RODs.
\subsection{STEP 2 : Construction of  new sets of rate-1 RODs}
\label{subsec2-2}
Transition from a square ROD to rate-1 ROD is illustrated in \cite{TJC} using column vector representation of a ROD. In a similar way, we construct a rate-1 ROD $W_n$ of size $[\nu(n),n,\nu(n)]$ for $n$ transmit antennas from a ROD of size $[\nu(n),\nu(n),n]$ where $n$ is any non-zero positive integer, not necessarily power of 2. 

Any square ROD of order $\nu(n)$ obtained via a suitable pair of mapping $\gamma_{\nu(n)}$ and $\psi_{\nu(n)}$ satisfying the conditions of Theorem \ref{ratesquarerod} (for instance, $R_{\nu(n)}$ obtained in the previous subsection or  $A_{\nu(n)},\hat{A}_{\nu(n)}$ and $G_{\nu(n)}$ obtained in Appendix \ref{appendixII}) can be used for this purpose. We refer to any such design by $B_{\nu(n)}$ consisting of $n$ real variables $z_0,z_1,\cdots,z_{n-1}$.

Let $y_0,y_1,\cdots,y_{\nu(n)-1}$ be $\nu(n)$ real variables which constitute the matrix $W_n$.
The matrix $W_n$ is obtained as follows: Make
$W_n(i,j)=0$ if the $i$-th row of $B_{\nu(n)}$  does not contain $z_j$. Otherwise,
$W_n(i,j)=y_k$ or $-y_k$ if $B_{\nu(n)}(i,k)=z_j$ or $-z_j$ respectively.
The construction of the matrix $W_n$ ensures that it is a rate-1 ROD. 
Using \eqref{definesf1},\eqref{definesf} and Theorem \ref{orthog}, we have
\begin{eqnarray}
\label{defineW}
 W_n(i,j)=s(i,j)y_{f(i,j)},
\end{eqnarray}
where 
\begin{eqnarray}
 s(i,j)&=&(-1)^{\vert i\cdot\psi_{\nu(n)}(\gamma_{\nu(n)}(j))\vert} \\
 f(i,j)&=&i\oplus \gamma_{\nu(n)}(j) \nonumber
 \end{eqnarray}
for $0\leq i\leq \nu(n)-1,~ 0\leq j\leq n-1$.\\
Let $\hat{W}_n$ be a matrix  which is similar to the matrix $W_n$, given by
\begin{eqnarray}
\label{defineWhat}
 \hat{W}_n(i,j)=\hat{s}(i,j)y_{f(i,j)},
\end{eqnarray}
where 
\begin{eqnarray}
 \hat{s}(i,j)&=&(-1)^{\vert (i\oplus \gamma_{\nu(n)}(j))\cdot\psi_{\nu(n)}(\gamma_{\nu(n)}(j))\vert}\\
 f(i,j)&=&i\oplus \gamma_{\nu(n)}(j) \nonumber.
\end{eqnarray}
$\hat{W}_n$ is also a rate-1 ROD. Both $W_n$ and $\hat{W}_n$ are the new sets of rate-1 RODs that are used in the following subsection to construct our codes.

As examples, the RODs $W_9$ and $\hat{W}_9$ of size $[16,9,16]$ obtained using $R_{16}$ are given by \eqref{W9} and \eqref{What9} respectively and the RODs $W_{10}$ and $\hat{W}_{10}$ of size $[32,10,32]$ obtained using $R_{32}$ are given by \eqref{W10} and \eqref{What10} respectively.

{\footnotesize
\begin{equation}
\label{W9}
\hspace{-10pt}
W_9=\left[ \hspace{-5pt}
\begin{array}{  r @{\hspace{.2pt}} r @{\hspace{.2pt}} r @{\hspace{.2pt}} r @{\hspace{.2pt}} r @{\hspace{.2pt}} r @{\hspace{.2pt}} r @{\hspace{.2pt}} r @{\hspace{.2pt}} r}
   y_0 &  y_1 &  y_2 &  y_3 &  y_4 &  y_5 &  y_6 &  y_7 &  y_8 \\
   y_1 & -y_0 &  y_3 & -y_2 &  y_5 & -y_4 & -y_7 &  y_6 &  y_9 \\
   y_2 & -y_3 & -y_0 &  y_1 &  y_6 &  y_7 & -y_4 & -y_5 & y_{10} \\
   y_3 &  y_2 & -y_1 & -y_0 &  y_7 & -y_6 &  y_5 & -y_4 & y_{11} \\
   y_4 & -y_5 & -y_6 & -y_7 & -y_0 &  y_1 &  y_2 &  y_3 & y_{12} \\
   y_5 &  y_4 & -y_7 &  y_6 & -y_1 & -y_0 & -y_3 &  y_2 & y_{13} \\
   y_6 &  y_7 &  y_4 & -y_5 & -y_2 &  y_3 & -y_0 & -y_1 & y_{14} \\
   y_7 & -y_6 &  y_5 &  y_4 & -y_3 & -y_2 &  y_1 & -y_0 & y_{15} \\
   y_8 & -y_9 & -y_{10} & -y_{11} & -y_{12} & -y_{13} & -y_{14} & -y_{15} & -y_0 \\
   y_9 &  y_8 & -y_{11} & y_{10} & -y_{13} & y_{12} & y_{15} & -y_{14} & -y_1 \\
  y_{10} & y_{11} &  y_8 & -y_9 & -y_{14} & -y_{15} & y_{12} & y_{13} & -y_2 \\
  y_{11} & -y_{10}&  y_9 &  y_8 & -y_{15} & y_{14} & -y_{13} & y_{12} & -y_3 \\
  y_{12} & y_{13} & y_{14} & y_{15} &  y_8 & -y_9 & -y_{10} & -y_{11} & -y_4 \\
  y_{13} & -y_{12}& y_{15} & -y_{14} &  y_9 &  y_8 & y_{11} & -y_{10} & -y_5 \\
  y_{14} & -y_{15}& -y_{12} & y_{13} & y_{10} & -y_{11} &  y_8 &  y_9 & -y_6 \\
  y_{15} & y_{14} & -y_{13} & -y_{12} & y_{11} & y_{10} & -y_9 &  y_8 & -y_7
\end{array}\right]
\end{equation}
}
{\footnotesize
\begin{equation}
\label{What9}
\hspace{-10pt}
\hat{W}_9=\left[ \hspace{-5pt}
\begin{array}{ r @{\hspace{.2pt}} r @{\hspace{.2pt}} r @{\hspace{.2pt}} r @{\hspace{.2pt}} r @{\hspace{.2pt}} r @{\hspace{.2pt}} r @{\hspace{.2pt}} r @{\hspace{.2pt}} r}
   y_0 &    -y_1 &    -y_2 &     y_3 &    -y_4 &     y_5 &     y_6 &    -y_7 &    -y_8\\
   y_1 &     y_0 &    -y_3 &    -y_2 &    -y_5 &    -y_4 &    -y_7 &    -y_6 &    -y_9\\
   y_2 &     y_3 &     y_0 &     y_1 &    -y_6 &     y_7 &    -y_4 &     y_5 &   -y_{10}\\
   y_3 &    -y_2 &     y_1 &    -y_0 &    -y_7 &    -y_6 &     y_5 &     y_4 &   -y_{11}\\
   y_4 &     y_5 &     y_6 &    -y_7 &     y_0 &     y_1 &     y_2 &    -y_3 &   -y_{12}\\
   y_5 &    -y_4 &     y_7 &     y_6 &     y_1 &    -y_0 &    -y_3 &    -y_2 &   -y_{13}\\
   y_6 &    -y_7 &    -y_4 &    -y_5 &     y_2 &     y_3 &    -y_0 &     y_1 &   -y_{14}\\
   y_7 &     y_6 &    -y_5 &     y_4 &     y_3 &    -y_2 &     y_1 &     y_0 &   -y_{15}\\
   y_8 &     y_9 &    y_{10} &   -y_{11} &    y_{12} &   -y_{13} &   -y_{14} &    y_{15} &    y_0\\
   y_9 &    -y_8 &    y_{11} &    y_{10} &    y_{13} &    y_{12} &    y_{15} &    y_{14} &    y_1\\
  y_{10} &   -y_{11} &    -y_8 &    -y_9 &    y_{14} &   -y_{15} &    y_{12} &   -y_{13} &    y_2\\
  y_{11} &    y_{10} &    -y_9 &     y_8 &    y_{15} &    y_{14} &   -y_{13} &   -y_{12} &    y_3\\
  y_{12} &   -y_{13} &   -y_{14} &    y_{15} &    -y_8 &    -y_9 &   -y_{10} &    y_{11} &    y_4\\
  y_{13} &    y_{12} &   -y_{15} &   -y_{14} &    -y_9 &     y_8 &    y_{11} &    y_{10} &    y_5\\
  y_{14} &    y_{15} &    y_{12} &    y_{13} &   -y_{10} &   -y_{11} &     y_8 &    -y_9 &    y_6\\
  y_{15} &   -y_{14} &    y_{13} &   -y_{12} &   -y_{11} &    y_{10} &    -y_9 &    -y_8 &    y_7\\
\end{array}\right]
\end{equation}
}
{\footnotesize
\begin{equation}
\label{W10}
\hspace{-10pt}
W_{10}=\left[ \hspace{-5pt}
\begin{array}{ r @{\hspace{.2pt}} r @{\hspace{.2pt}} r @{\hspace{.2pt}} r @{\hspace{.2pt}} r @{\hspace{.2pt}} r @{\hspace{.2pt}} r @{\hspace{.2pt}} r @{\hspace{.2pt}} r @{\hspace{.2pt}} r}
   y_0 &  y_1 &  y_2 &  y_3 &  y_4 &  y_5 &  y_6 &  y_7 &  y_8 & y_{16} \\
   y_1 & -y_0 &  y_3 & -y_2 &  y_5 & -y_4 & -y_7 &  y_6 &  y_9 & y_{17} \\
   y_2 & -y_3 & -y_0 &  y_1 &  y_6 &  y_7 & -y_4 & -y_5 & y_{10} & y_{18} \\
   y_3 &  y_2 & -y_1 & -y_0 &  y_7 & -y_6 &  y_5 & -y_4 & y_{11} & y_{19} \\
   y_4 & -y_5 & -y_6 & -y_7 & -y_0 &  y_1 &  y_2 &  y_3 & y_{12} & y_{20} \\
   y_5 &  y_4 & -y_7 &  y_6 & -y_1 & -y_0 & -y_3 &  y_2 & y_{13} & y_{21} \\
   y_6 &  y_7 &  y_4 & -y_5 & -y_2 &  y_3 & -y_0 & -y_1 & y_{14} & y_{22} \\
   y_7 & -y_6 &  y_5 &  y_4 & -y_3 & -y_2 &  y_1 & -y_0 & y_{15} & y_{23} \\
   y_8 & -y_9 & -y_{10} & -y_{11} & -y_{12} & -y_{13} & -y_{14} & -y_{15} & -y_0 & y_{24} \\
   y_9 &  y_8 & -y_{11} & y_{10} & -y_{13} & y_{12} & y_{15} & -y_{14} & -y_1 & y_{25} \\
  y_{10} & y_{11} &  y_8 & -y_9 & -y_{14} & -y_{15} & y_{12} & y_{13} & -y_2 & y_{26} \\
  y_{11} & -y_{10} &  y_9 &  y_8 & -y_{15} & y_{14} & -y_{13} & y_{12} & -y_3 & y_{27} \\
  y_{12} & y_{13} & y_{14} & y_{15} &  y_8 & -y_9 & -y_{10} & -y_{11} & -y_4 & y_{28} \\
  y_{13} & -y_{12} & y_{15} & -y_{14} &  y_9 &  y_8 & y_{11} & -y_{10} & -y_5 & y_{29} \\
  y_{14} & -y_{15} & -y_{12} & y_{13} & y_{10} & -y_{11} &  y_8 &  y_9 & -y_6 & y_{30} \\
  y_{15} & y_{14} & -y_{13} & -y_{12} & y_{11} & y_{10} & -y_9 &  y_8 & -y_7 & y_{31} \\
  y_{16} & -y_{17} & -y_{18} & -y_{19} & -y_{20} & -y_{21} & -y_{22} & -y_{23}& -y_{24}&-y_0\\
  y_{17} & y_{16} & -y_{19} & y_{18} & -y_{21} & y_{20} & y_{23} & -y_{22} & -y_{25} & -y_1 \\
  y_{18} & y_{19} & y_{16} & -y_{17} & -y_{22} & -y_{23} & y_{20} & y_{21} & -y_{26} & -y_2 \\
  y_{19} & -y_{18} & y_{17} & y_{16} & -y_{23} & y_{22} & -y_{21} & y_{20} & -y_{27} & -y_3 \\
  y_{20} & y_{21} & y_{22} & y_{23} & y_{16} & -y_{17} & -y_{18} & -y_{19} & -y_{28} & -y_4 \\
  y_{21} & -y_{20} & y_{23} & -y_{22} & y_{17} & y_{16} & y_{19} & -y_{18} & -y_{29} & -y_5 \\
  y_{22} & -y_{23} & -y_{20} & y_{21} & y_{18} & -y_{19} & y_{16} & y_{17} & -y_{30} & -y_6 \\
  y_{23} & y_{22} & -y_{21} & -y_{20} & y_{19} & y_{18} & -y_{17} & y_{16} & -y_{31} & -y_7 \\
  y_{24} & y_{25} & y_{26} & y_{27} & y_{28} & y_{29} & y_{30} & y_{31} & y_{16} & -y_8 \\
  y_{25} & -y_{24} & y_{27} & -y_{26} & y_{29} & -y_{28} & -y_{31} & y_{30} & y_{17} & -y_9 \\
  y_{26} & -y_{27} & -y_{24} & y_{25} & y_{30} & y_{31} & -y_{28} & -y_{29}& y_{18}&-y_{10} \\
  y_{27} & y_{26} & -y_{25} & -y_{24} & y_{31} & -y_{30} & y_{29}&-y_{28} & y_{19}& -y_{11} \\
  y_{28} & -y_{29} & -y_{30} & -y_{31} & -y_{24} & y_{25} & y_{26} & y_{27} & y_{20}&-y_{12}\\
  y_{29} & y_{28} & -y_{31} & y_{30} & -y_{25} & -y_{24} & -y_{27} & y_{26}& y_{21}&-y_{13} \\
  y_{30} & y_{31} & y_{28} & -y_{29} & -y_{26} & y_{27} & -y_{24} & -y_{25}& y_{22}&-y_{14} \\
  y_{31} & -y_{30} & y_{29} & y_{28} & -y_{27} & -y_{26} & y_{25} & -y_{24} & y_{23} & -y_{15}
\end{array}\right]
\end{equation}
}
\vspace{-0.6cm}
{\footnotesize
\begin{equation}
\label{What10}
\hspace{-10pt}
\hat{W}_{10}=\left[ \hspace{-5pt}
\begin{array}{ r @{\hspace{.2pt}} r @{\hspace{.2pt}} r @{\hspace{.2pt}} r @{\hspace{.2pt}} r @{\hspace{.2pt}} r @{\hspace{.2pt}} r @{\hspace{.2pt}} r @{\hspace{.2pt}} r @{\hspace{.2pt}} r}
   y_0 &    -y_1 &    -y_2 &     y_3 &    -y_4 &     y_5 &     y_6 &    -y_7 &    -y_8 &   -y_{16}\\
   y_1 &     y_0 &    -y_3 &    -y_2 &    -y_5 &    -y_4 &    -y_7 &    -y_6 &    -y_9 &   -y_{17}\\
   y_2 &     y_3 &     y_0 &     y_1 &    -y_6 &     y_7 &    -y_4 &     y_5 &   -y_{10} &  -y_{18}\\
   y_3 &    -y_2 &     y_1 &    -y_0 &    -y_7 &    -y_6 &     y_5 &     y_4 &   -y_{11} &  -y_{19}\\
   y_4 &     y_5 &     y_6 &    -y_7 &     y_0 &     y_1 &     y_2 &    -y_3 &   -y_{12} &  -y_{20}\\
   y_5 &    -y_4 &     y_7 &     y_6 &     y_1 &    -y_0 &    -y_3 &    -y_2 &   -y_{13} &  -y_{21}\\
   y_6 &    -y_7 &    -y_4 &    -y_5 &     y_2 &     y_3 &    -y_0 &     y_1 &   -y_{14} &  -y_{22}\\
   y_7 &     y_6 &    -y_5 &     y_4 &     y_3 &    -y_2 &     y_1 &     y_0 &   -y_{15} &  -y_{23}\\
   y_8 &     y_9 &    y_{10} &   -y_{11} &    y_{12} &   -y_{13} &   -y_{14} &    y_{15} &    y_0 &   -y_{24}\\
   y_9 &    -y_8 &    y_{11} &    y_{10} &    y_{13} &    y_{12} &    y_{15} &    y_{14} &    y_1 &   -y_{25}\\
  y_{10} &   -y_{11} &    -y_8 &    -y_9 &    y_{14} &   -y_{15} &    y_{12} &   -y_{13} &    y_2 &   -y_{26}\\
  y_{11} &    y_{10} &    -y_9 &     y_8 &    y_{15} &    y_{14} &   -y_{13} &   -y_{12} &    y_3 &   -y_{27}\\
  y_{12} &   -y_{13} &   -y_{14} &    y_{15} &    -y_8 &    -y_9 &   -y_{10} &    y_{11} &    y_4 &   -y_{28}\\
  y_{13} &    y_{12} &   -y_{15} &   -y_{14} &    -y_9 &     y_8 &    y_{11} &    y_{10} &    y_5 &   -y_{29}\\
  y_{14} &    y_{15} &    y_{12} &    y_{13} &   -y_{10} &   -y_{11} &     y_8 &    -y_9 &    y_6 &   -y_{30}\\
  y_{15} &   -y_{14} &    y_{13} &   -y_{12} &   -y_{11} &    y_{10} &    -y_9 &    -y_8 &    y_7 &   -y_{31}\\
  y_{16} &    y_{17} &    y_{18} &   -y_{19} &    y_{20} &   -y_{21} &   -y_{22} &    y_{23} &    y_{24} &     y_0\\
  y_{17} &   -y_{16} &    y_{19} &    y_{18} &    y_{21} &    y_{20} &    y_{23} &    y_{22} &    y_{25} &     y_1\\
  y_{18} &   -y_{19} &   -y_{16} &   -y_{17} &    y_{22} &   -y_{23} &    y_{20} &   -y_{21} &    y_{26} &     y_2\\
  y_{19} &    y_{18} &   -y_{17} &    y_{16} &    y_{23} &    y_{22} &   -y_{21} &   -y_{20} &    y_{27} &     y_3\\
  y_{20} &   -y_{21} &   -y_{22} &    y_{23} &   -y_{16} &   -y_{17} &   -y_{18} &    y_{19} &    y_{28} &     y_4\\
  y_{21} &    y_{20} &   -y_{23} &   -y_{22} &   -y_{17} &    y_{16} &    y_{19} &    y_{18} &    y_{29} &     y_5\\
  y_{22} &    y_{23} &    y_{20} &    y_{21} &   -y_{18} &   -y_{19} &    y_{16} &   -y_{17} &    y_{30} &     y_6\\
  y_{23} &   -y_{22} &    y_{21} &   -y_{20} &   -y_{19} &    y_{18} &   -y_{17} &   -y_{16} &    y_{31} &     y_7\\
  y_{24} &   -y_{25} &   -y_{26} &    y_{27} &   -y_{28} &    y_{29} &    y_{30} &   -y_{31} &   -y_{16} &     y_8\\
  y_{25} &    y_{24} &   -y_{27} &   -y_{26} &   -y_{29} &   -y_{28} &   -y_{31} &   -y_{30} &   -y_{17} &     y_9\\
  y_{26} &    y_{27} &    y_{24} &    y_{25} &   -y_{30} &    y_{31} &   -y_{28} &    y_{29} &   -y_{18} &    y_{10}\\
  y_{27} &   -y_{26} &    y_{25} &   -y_{24} &   -y_{31} &   -y_{30} &    y_{29} &    y_{28} &   -y_{19} &    y_{11}\\
  y_{28} &    y_{29} &    y_{30} &   -y_{31} &    y_{24} &    y_{25} &    y_{26} &   -y_{27} &   -y_{20} &    y_{12}\\
  y_{29} &   -y_{28} &    y_{31} &    y_{30} &    y_{25} &   -y_{24} &   -y_{27} &   -y_{26} &   -y_{21} &    y_{13}\\
  y_{30} &   -y_{31} &   -y_{28} &   -y_{29} &    y_{26} &    y_{27} &   -y_{24} &    y_{25} &   -y_{22} &    y_{14}\\
  y_{31} &    y_{30} &   -y_{29} &    y_{28} &    y_{27} &   -y_{26} &    y_{25} &    y_{24} &   -y_{23} &    y_{15}\\
\end{array}\right]
 \end{equation}
}
\subsection{STEP 3 : Construction of low-delay rate-1/2 Scaled CODs}
\label{subsec2-3}
In this subsection, we construct a rate-1/2 scaled COD with the help of rate-1 RODs $W_n$ and $\hat{W}_n$  constructed in the previous subsection.

Let $x_0,x_1,\cdots$ be complex variables. The $8\times 8,$ rate-$\frac{1}{2}$ CODs $A(x_0,x_1,x_2,x_3)$ and $B(x_4,x_5,x_6,x_7)$ and the $8\times 1$ column vector $C(x_0,x_1,x_2,x_3)$ in four variables, shown below, are the basic ingredients for our construction of rate-$\frac{1}{2}$ scaled CODs. 

{\footnotesize
\begin{equation}
\label{tx8cod1}
   A(x_0,x_1,x_1,x_3)=\left[
\begin{array}{ r @{\hspace{.2pt}} r @{\hspace{.2pt}} r @{\hspace{.2pt}} r @{\hspace{.2pt}} r @{\hspace{.2pt}} r @{\hspace{.2pt}} r @{\hspace{.2pt}} r @{\hspace{.2pt}} }
     x_0 & -x_1^* & -x_2^* & 0   & -x_3^* & 0 & 0 & 0 \\
     x_1 & x_0^*  & 0  & -x_2^* & 0 & -x_3^* & 0 & 0  \\
     x_2 & 0 & x_0^* & x_1^*  & 0 & 0 & -x_3^* & 0 \\
     0   & x_2 & -x_1 & x_0  & 0 & 0 & 0 & -x_3^*  \\
     x_3 & 0 & 0 & 0 & x_0^* & x_1^* &x_2^* &0    \\
     0   & x_3 & 0 & 0  & -x_1 & x_0 &0 &x_2^*    \\
     0   &0  & x_3 & 0  & -x_2 &0 & x_0 & -x_1^* \\
     0   &0  &0 & x_3 & 0  & -x_2  & x_1 & x_0^* 
\end{array}\right]
\end{equation}
}
{\footnotesize
\begin{equation}
   B(x_4,x_5,x_6,x_7)=\left[
\begin{array}{ r @{\hspace{.2pt}} r @{\hspace{.2pt}} r @{\hspace{.2pt}} r @{\hspace{.2pt}} r @{\hspace{.2pt}} r @{\hspace{.2pt}} r @{\hspace{.2pt}} r @{\hspace{.2pt}} }
    x_4 &-x_5^* &-x_6^* &-x_7^*  & 0 & 0 & 0 & 0 \\
    x_5 & x_4^* & 0 & 0  & -x_6^* &-x_7^* & 0 & 0 \\
    x_6 &0  & x_4^* & 0  & x_5^* &0 &-x_7^* & 0 \\
     0  & x_6 & -x_5 &0  & x_4  & 0  & 0 &-x_7^* \\
     x_7&0 &0 & x_4^*  &0 & x_5^*   & x_6^*  & 0 \\
     0  & x_7  & 0  & -x_5 & 0 & x_4 & 0 & x_6^* \\
     0  & 0 & x_7 &-x_6  & 0 & 0 & x_4 &-x_5^* \\
     0  & 0 & 0 & 0 & x_7 &-x_6 & x_5   & x_4^*
    \end{array}\right]
\end{equation}
}
{\footnotesize
\begin{equation}
C(x_0,x_1,x_2,x_3) \\ 
=\frac{1}{\sqrt{2}}\left[
\begin{array}{ r @{\hspace{.2pt}} r @{\hspace{.2pt}} r @{\hspace{.2pt}} r @{\hspace{.2pt}} r @{\hspace{.2pt}} r @{\hspace{.2pt}} r @{\hspace{.2pt}} r @{\hspace{.2pt}} }
-x_3^* & x_2^* &-x_1^* &-x_0 & x_0^* &-x_1 &-x_2  &-x_3
\end{array}\right]^\mathcal{T}.
\end{equation}
}
Let $i$ be any non-negative integer. Define the following matrices:
\begin{eqnarray}
A(2i)&=&A(x_{8i},x_{8i+1},x_{8i+2},x_{8i+3}),\nonumber\\
A(2i+1)&=&B(x_{8i+4},x_{8i+5},x_{8i+6},x_{8i+7}),\\ \overline{A}(i)&=&C(x_{4i},x_{4i+1},x_{4i+2},x_{4i+3})\nonumber.
\end{eqnarray}

One can easily verify that the matrices given by
\begin{equation}
\label{a0a1}
  \left[\begin{array}{cc}
    A(0) &\overline{A}(1)\\
    A(1) &\overline{A}(0)
   \end{array}\right],~~
\left[\begin{array}{cc}
    \overline{A}(1) &-\overline{A}(0)\\
    \overline{A}(0) &\overline{A}(1)
   \end{array}\right],~~
\end{equation}
are scaled CODs as the columns of the matrices are orthogonal to each other and 
the norm of each column  is equal to square root of the sum of the norms of the variables of the design.\\
By relabeling the variables in the matrices $A(0),A(1),\overline{A}(0)$ and  $\overline{A}(1)$, it follows that
\begin{equation}
\label{Aij}
  \left[\begin{array}{cc}
    A(i) &\overline{A}(j)\\
    A(j) &\overline{A}(i)
   \end{array}\right]
\end{equation}
is a scaled COD whenever $(i+j)$ is odd and
\begin{equation}
\label{Aoverline}
\left[\begin{array}{cc}
    \overline{A}(i) &-\overline{A}(j)\\
    \overline{A}(j) &\overline{A}(i)
   \end{array}\right],~~
\end{equation}
is a scaled COD for all values of $i$ and $j$, $i\neq j$.

Let $n$ be an integer such that $n\geq 9$.  We construct a matrix $(DR)_n$ of size $\nu(n)\times n$ as follows:
Let $t=n-8$ and  $W_t$ and $\hat{W}_t$ are the two rate-1 RODs of size $[\nu(t),t,\nu(t)]$ in $\nu(t)$ real variables $y_0,y_1,\cdots,y_{\nu(t)-1}$ constructed in the previous subsection. Let $H_t$ and $\hat{H}_t$ be the matrices formed by substituting $y_i$ with $\overline{A}(2i+1)$ in the matrix $W_t$ and $\overline{A}(2i)$ in the matrix $\hat{W}_t$ respectively for $i=0$ to $\nu(t)-1$.
Note that the size of both $H_t$ and $\hat{H}_t$ is $8\nu(t)\times t$. 

Let $u=\nu(n)/8$. Let $E_8$ and $O_8$, each of size $4u\times 8$, be defined as follows:
\begin{eqnarray}
\label{rowvectorrep}
E_8= 
\left[
\begin{array}{c}
A(0) \\
A(2) \\
. \\
. \\
. \\
A(u-2)
\end{array}
\right]~~
O_8=
\left[
\begin{array}{c}
A(1) \\
A(3) \\
. \\
. \\
. \\
A(u-1)
\end{array}
\right].
\end{eqnarray}
Define the matrix $(DR)_n$ as
\begin{equation}
\label{gn}
   (DR)_n=\left[\begin{array}{cc}
     E_8 & H_t\\
     O_8 & \hat{H}_t 
\end{array}\right].
\end{equation}
Note that the number of rows and columns of the matrix $(DR)_n$ are $16\cdot \nu(n-8)=8\cdot \nu(n)/8 =\nu(n)$ and $t+8=n$ respectively.
The following theorem is the main result of this paper.
\begin{theorem}
\label{rate12cod}
Let $n$ be a positive integer and $(DR)_n$ be the matrix as defined in \eqref{gn}.
Then $(DR)_n$  is a rate-1/2 scaled COD of size $[\nu(n),n,\frac{\nu(n)}{2}]$.
\end{theorem} 
\begin{proof}
For $n\leq 8$, one can construct rate-1/2 COD of size $[\nu(n),n,\frac{\nu(n)}{2}]$
from  a COD of size $[8,8,4]$ given in \eqref{tx8cod1}. We assume that $n\geq 9$.
Let $p=\nu(n)$.
We have
\begin{equation*}
(DR)_n^H(DR)_n=\left[\begin{array}{cc}
     E_8^HE_8+O_8^HO_8  & E_8^HH_t + O_8^H\hat{H}_t\\
     H_t^HE_8 + \hat{H}_t^HO_8 & H_t^HH_t+\hat{H}_t^H\hat{H}_t
\end{array}\right].
\end{equation*}
From the construction of $E_8$ and $O_8$ given in \eqref{rowvectorrep},
we have \\
 $E_8^HE_8+O_8^HO_8= ({\vert x_0\vert}^2 +\cdots+{\vert x_{\frac{p}{2}-1}\vert}^2)I_8$.

From equation \eqref{Aoverline}, we have
\begin{eqnarray*}
H_t^HH_t+\hat{H}_t^H\hat{H}_t= ({\vert x_0\vert}^2 +\cdots+{\vert x_{p/2-1}\vert}^2)I_{n-8},\\
\end{eqnarray*}

Thus it is enough to show that $E_8^HH_t + O_8^H\hat{H}_t=0_{8\times (n-8)}$ where $0_{8\times (n-8)}$ is a matrix of size $8\times (n-8)$ containing zero only.
Let the $j$-th column of $H_t$ and $\hat{H}_t$ be $H_t(j)$ and $\hat{H}_t(j)$ respectively.
Then we show that $Z(j)=E_8^HH_t(j) + O_8^H\hat{H}_t(j)=0_{8\times 1}$ for all $j\in\{0,1,\cdots,n-8-1\}$.\\
Let $u=p/8$. For convenience, we write $\gamma$ for $\gamma_{\nu(t)}$.
We have
\begin{eqnarray*}
\begin{array}{c}
E_8^H= \left[
\begin{array}{cccc}
A^H(0) & A^H(2)&\cdots & A^H(u-2)
\end{array}
\right],\\
\vspace*{.40cm}
O_8^H= \left[
\begin{array}{cccc}
A^H(1) & A^H(3)&\cdots & A^H(u-1)
\end{array}
\right],\\
\vspace*{.40cm}
H_t(j)=
\left[
\begin{array}{c}
s(0,j)\overline{A}({2(0\oplus \gamma(j))+1}) \\
s(1,j)\overline{A}({2(1\oplus \gamma(j))+1}) \\
. \\
. \\
s(i,j)\overline{A}({2(i\oplus \gamma(j))+1}) \\
. \\
. \\
s(\frac{u}{2}-1,j)\overline{A}({2\bigl((\frac{u}{2}-1)\oplus \gamma(j)\bigr)+1})
\end{array}
\right], \\
\vspace*{.40cm}
\hat{H}_t(j)=
\left[
\begin{array}{c}
\hat{s}(0,j)\overline{A}({2(0\oplus \gamma(j))}) \\
\hat{s}(1,j)\overline{A}({2(1\oplus \gamma(j))}) \\
. \\
. \\
\hat{s}(i,j)\overline{A}({2(i\oplus \gamma(j))}) \\
. \\
. \\
\hat{s}(\frac{u}{2}-1,j)\overline{A}({2((\frac{u}{2}-1)\oplus \gamma(j))})
\end{array}
\right],
\end{array}
\end{eqnarray*}
where $s(i,j)$ and $\hat{s}(i,j)$ are defined in \eqref{defineW} and \eqref{defineWhat} respectively. We have
\begin{eqnarray*}
Z(j)&=&\sum_{i=0}^{\frac{u}{2}-1}s(i,j)A^H(2i)\overline{A}({2(i\oplus \gamma(j))+1})\\
 &&+ \sum_{i=0}^{\frac{u}{2}-1}\hat{s}(i,j)A^H(2i+1)\overline{A}({2(i\oplus \gamma(j))}).
\end{eqnarray*}
Now
\begin{eqnarray*}
\sum_{i=0}^{\frac{u}{2}-1}\hat{s}(i,j)A^H(2i+1)\overline{A}({2(i\oplus \gamma(j))})\\
=\sum_{i=0}^{\frac{u}{2}-1}\hat{s}(i\oplus \gamma(j),j)A^H(2(i\oplus \gamma(j))+1)\overline{A}({2i})\bigr)\\
\mbox {and } s(i,j)=\hat{s}(i\oplus \gamma(j),j).
\end{eqnarray*}
Hence
\begin{eqnarray*}
Z(j)&=&\sum_{i=0}^{\frac{u}{2}-1}\bigl(s(i,j)A^H(2i)\overline{A}({2(i\oplus \gamma(j))+1})\\  &&+ \hat{s}(i\oplus \gamma(j),j)A^H(2(i\oplus \gamma(j))+1)\overline{A}({2i})\bigr)\\
&=&\sum_{i=0}^{\frac{u}{2}-1}s(i,j)(A^H(2i)\overline{A}({2(i\oplus \gamma(j))+1})\\
  &&+ A^H(2(i\oplus \gamma(j))+1)\overline{A}({2i}))\\
 &=&0_{8 \times 1} \mbox{ using \eqref{Aij}.}
\end{eqnarray*}
\end{proof}

We illustrate our main result in the following example. 
\begin{example}
For $9$ transmit antennas, we have rate-$1/2$ scaled COD of size $[16,9,8]$ given by
{\tiny
\begin{equation}
   \left[\begin{array}{ccccccccc}
     x_0 & -x_1^* & -x_2^* & 0   & -x_3^* & 0 & 0 & 0 &\frac{-x_7^*}{\sqrt{2}} \\
     x_1 & x_0^*  & 0  & -x_2^* & 0 & -x_3^* & 0 & 0  &\frac{x_6^*}{\sqrt{2}}  \\
     x_2 & 0 & x_0^* & x_1^*  & 0 & 0 & -x_3^* & 0 &\frac{-x_5^*}{\sqrt{2}} \\
     0   & x_2 & -x_1 & x_0  & 0 & 0 & 0 & -x_3^*  &\frac{-x_4}{\sqrt{2}} \\
     x_3 & 0 & 0 & 0 & x_0^* & x_1^* &x_2^* &0    &\frac{x_4^*}{\sqrt{2}}   \\
     0   & x_3 & 0 & 0  & -x_1 & x_0 &0 &x_2^*     &\frac{-x_5}{\sqrt{2}} \\
     0   &0  & x_3 & 0  & -x_2 &0 & x_0 & -x_1^* &\frac{-x_6}{\sqrt{2}} \\
     0   &0  &0 & x_3 & 0  & -x_2  & x_1 & x_0^* &\frac{-x_7}{\sqrt{2}} \\
     x_4 &-x_5^* &-x_6^* &-x_7^*  & 0 & 0 & 0 & 0  &\frac{-x_3^*}{\sqrt{2}}\\
     x_5& x_4^* & 0 & 0  & -x_6^* &-x_7^* & 0 & 0   &\frac{x_2^*}{\sqrt{2}} \\
    x_6 &0  & x_4^* & 0  & x_5^* &0 &-x_7^* & 0 &\frac{-x_1^*}{\sqrt{2}}  \\
     0    & x_6 & -x_5 &0  & x_4  & 0  & 0 &-x_7^* &\frac{-x_0}{\sqrt{2}}\\
     x_7 &0 &0 & x_4^*  &0 & x_5^*   &-x_7^*  & 0  &\frac{x_0^*}{\sqrt{2}} \\
     0    & x_7  & 0  & -x_5 & 0 & x_4 & 0 & x_6^*  &\frac{-x_1}{\sqrt{2}}  \\
     0    & 0 & x_7 &-x_6  & 0 & 0 & x_4 &-x_5^* &\frac{-x_2}{\sqrt{2}}   \\
     0    & 0 & 0 & 0 & x_7 &-x_6 & x_5   & x_4^*  &\frac{-x_3}{\sqrt{2}}\\
\end{array}\right].
 \end{equation}
 }
while the known rate $1/2$ scaled COD for $9$ transmit antenna is given by 
\cite{TJC}
{ \tiny \[  
\frac{1}{\sqrt{2}}\cdot
\left[\begin{array}{c@{\hspace{0.2cm}}c@{\hspace{0.2cm}}c@{\hspace{0.2cm}}c@{\hspace{0.2cm}}c@{\hspace{0.2cm}}c@{\hspace{0.2cm}}c@{\hspace{0.2cm}}c@{\hspace{0.2cm}}c@{\hspace{0.2cm}}}
    x_0   &-x_1   &-x_2    &-x_3    &-x_4    &-x_5     &-x_6  &-x_7 &-x_8\\
    x_1   & x_0   & x_3    &-x_2    & x_5    &-x_4     &-x_7  & x_6 & x_9\\
    x_2   &-x_3   & x_0    & x_1    & x_6    &x_7      &-x_4  &-x_5 & x_{10}\\
    x_3   & x_2   &-x_1    & x_0    & x_7    &-x_6     & x_5  &-x_4 & x_{11}\\
    x_4   &-x_5   &-x_6    &-x_7    & x_0    & x_1     & x_2  & x_3 & x_{12}\\
    x_5   & x_4   &-x_7    & x_6    &-x_1    & x_0     &-x_3  & x_2 & x_{13}  \\
    x_6   & x_7   & x_4    &-x_5    &-x_2    &x_3      & x_0  &-x_1 & x_{14}  \\
    x_7   &-x_6   & x_5    & x_4    &-x_3    &-x_2     & x_1  & x_0 & x_{15}\\
    x_8&-x_9&-x_{10} &-x_{11} &-x_{12} &-x_{13}  &-x_{14}&-x_{15} & x_0\\
    x_9& x_8   &-x_{11} & x_{10} &-x_{13} & x_{12}  & x_{15}&-x_{14}&-x_1\\
    x_{10}& x_{11}& x_8    &-x_9 &-x_{14} &-x_{15}  & x_{12}& x_{13}&-x_2\\
    x_{11}&-x_{10}& x_9 & x_8    &-x_{15} & x_{14}  &-x_{13}& x_{12}&-x_3\\
    x_{12}& x_{13}& x_{14} & x_{15} & x_8    &-x_9  &-x_{10}&-x_{11}&-x_4\\
    x_{13}&-x_{12}& x_{15} &-x_{14} & x_9 & x_8     & x_{11}&-x_{10}&-x_5   \\
    x_{14}&-x_{15}&-x_{12} & x_{13} & x_{10} &-x_{11}  & x_8   & x_9&-x_6\\
    x_{15}&x_{14} &-x_{13} &-x_{12} & x_{11} & x_{10}  &-x_9& x_8   &-x_7\\
    x^*_0   &-x^*_1   &-x^*_2    &-x^*_3    &-x^*_4    &-x^*_5     &-x^*_6  &-x^*_7 &-x^*_8\\
    x^*_1   & x^*_0   & x^*_3    &-x^*_2    & x^*_5    &-x^*_4     &-x^*_7  & x^*_6 & x^*_9\\
    x^*_2   &-x^*_3   & x^*_0    & x^*_1    & x^*_6    &x^*_7      &-x^*_4  &-x^*_5 & x^*_{10}\\
    x^*_3   & x^*_2   &-x^*_1    & x^*_0    & x^*_7    &-x^*_6     & x^*_5  &-x^*_4 & x^*_{11}\\
    x^*_4   &-x^*_5   &-x^*_6    &-x^*_7    & x^*_0    & x^*_1     & x^*_2  & x^*_3 & x^*_{12}\\
    x^*_5   & x^*_4   &-x^*_7    & x^*_6    &-x^*_1    & x^*_0     &-x^*_3  & x^*_2 & x^*_{13}  \\
    x^*_6   & x^*_7   & x^*_4    &-x^*_5    &-x^*_2    &x^*_3      & x^*_0  &-x^*_1 & x^*_{14}  \\
    x^*_7   &-x^*_6   & x^*_5    & x^*_4    &-x^*_3    &-x^*_2     & x^*_1  & x^*_0 & x^*_{15}\\
    x^*_8&-x^*_9&-x^*_{10} &-x^*_{11} &-x^*_{12} &-x^*_{13}  &-x^*_{14}&-x^*_{15} & x^*_0\\
    x^*_9& x^*_8   &-x^*_{11} & x^*_{10} &-x^*_{13} & x^*_{12}  & x^*_{15}&-x^*_{14}&-x^*_1\\
    x^*_{10}& x^*_{11}& x^*_8    &-x^*_9 &-x^*_{14} &-x^*_{15}  & x^*_{12}& x^*_{13}&-x^*_2\\
    x^*_{11}&-x^*_{10}& x^*_9 & x^*_8    &-x^*_{15} & x^*_{14}  &-x^*_{13}& x^*_{12}&-x^*_3\\
    x^*_{12}& x^*_{13}& x^*_{14} & x^*_{15} & x^*_8    &-x^*_9  &-x^*_{10}&-x^*_{11}&-x^*_4\\
    x^*_{13}&-x^*_{12}& x^*_{15} &-x^*_{14} & x^*_9 & x^*_8     & x^*_{11}&-x^*_{10}&-x^*_5   \\
    x^*_{14}&-x^*_{15}&-x^*_{12} & x^*_{13} & x^*_{10} &-x^*_{11}  & x^*_8   & x^*_9&-x^*_6\\
    x^*_{15}&x^*_{14} &-x^*_{13} &-x^*_{12} & x^*_{11} & x^*_{10}  &-x^*_9& x^*_8   &-x^*_7
\end{array}\right].
\]}
where the decoding delay is $32$.
\end{example}
For $10$ transmit antennas, the rate-1/2 scaled COD given by Tarokh et al \cite{TJC} of size $[64,10,32]$ is given in Appendix \ref{appendixVI}, while
the new rate-$1/2$ code of size $[32,10,16]$ is given by \eqref{new321016}.

{\footnotesize
\begin{equation}
\label{new321016}
\hspace{-10pt}
\left[ \hspace{-5pt}
\begin{array}{ r @{\hspace{.2pt}} r @{\hspace{.2pt}} r @{\hspace{.2pt}} r @{\hspace{.2pt}} r @{\hspace{.2pt}} r @{\hspace{.2pt}} r @{\hspace{.2pt}} r @{\hspace{.2pt}} r @{\hspace{.2pt}} r}
     x_0 & -x_1^* & -x_2^* & 0   & -x_3^* & 0 & 0 & 0 &-\frac{x_7^*}{\sqrt{2}} & -\frac{x_{15}^*}{\sqrt{2}}\\
     x_1 & x_0^*  & 0  & -x_2^* & 0 & -x_3^* & 0 & 0  & \frac{x_6^*}{\sqrt{2}} & \frac{x_{14}^*}{\sqrt{2}}\\
     x_2 & 0 & x_0^* & x_1^*  & 0 & 0 & -x_3^* & 0 &-\frac{x_5^*}{\sqrt{2}} & -\frac{x_{13}^*}{\sqrt{2}}\\
     0   & x_2 & -x_1 & x_0  & 0 & 0 & 0 & -x_3^*  &-\frac{x_4}{\sqrt{2}} & -\frac{x_{12}}{\sqrt{2}}\\
     x_3 & 0 & 0 & 0 & x_0^* & x_1^* &x_2^* &0    & \frac{x_4^*}{\sqrt{2}}  & \frac{x_{12}^*}{\sqrt{2}} \\
     0   & x_3 & 0 & 0  & -x_1 & x_0 &0 &x_2^*     &-\frac{x_5}{\sqrt{2}} & -\frac{x_{13}}{\sqrt{2}}\\
     0   &0  & x_3 & 0  & -x_2 &0 & x_0 & -x_1^* &-\frac{x_6}{\sqrt{2}} & -\frac{x_{14}}{\sqrt{2}}\\
     0   &0  &0 & x_3 & 0  & -x_2  & x_1 & x_0^* &-\frac{x_7}{\sqrt{2}} & -\frac{x_{15}}{\sqrt{2}}\\
     x_8  & -x_9^* & -x_{10}^* & 0   & -x_{11}^* & 0 & 0 & 0 &-\frac{x_{15}^*}{\sqrt{2}} & \frac{x_7^*}{\sqrt{2}}\\
     x_9 & x_8^*  & 0  & -x_{10}^* & 0 & -x_{11}^* & 0 & 0  & \frac{x_{14}^*}{\sqrt{2}} &-\frac{x_6^*}{\sqrt{2}}\\
     x_{10} & 0 & x_8^* & x_9^*  & 0 & 0 & -x_{11}^* & 0 &-\frac{x_{13}^*}{\sqrt{2}} & \frac{x_5^*}{\sqrt{2}}\\
     0  & x_{10} & -x_9 & x_8  & 0 & 0 & 0 & -x_{11}^*  &-\frac{x_{12}}{\sqrt{2}} & \frac{x_4}{\sqrt{2}}\\
     x_{11} & 0 & 0 & 0 & x_8^* & x_9^* &x_{10}^* &0    & \frac{x_{12}^*}{\sqrt{2}} &-\frac{x_4^*}{\sqrt{2}}\\
     0  & x_{11} & 0 & 0  & -x_9 & x_8 &0 &x_{10}^*     &-\frac{x_{13}}{\sqrt{2}} & \frac{x_5}{\sqrt{2}} \\
     0  &0  & x_{11} & 0  & -x_{10} &0 & x_8 & -x_9^* &-\frac{x_{14}}{\sqrt{2}} & \frac{x_6}{\sqrt{2}}\\
     0  &0  &0 & x_{11} & 0  & -x_{10}  & x_9 & x_8^* &-\frac{x_{15}}{\sqrt{2}} & \frac{x_7}{\sqrt{2}} \\
    x_4 &-x_5^* &-x_6^* &-x_7^*  & 0 & 0 & 0 & 0  &-\frac{x_3^*}{\sqrt{2}} & \frac{x_{11}^*}{\sqrt{2}}\\
     x_5& x_4^* & 0 & 0  & -x_6^* &-x_7^* & 0 & 0   &\frac{x_2^*}{\sqrt{2}} &-\frac{x_{10}^*}{\sqrt{2}}\\
    x_6 &0  & x_4^* & 0  & x_5^* &0 &-x_7^* & 0 &-\frac{x_1^*}{\sqrt{2}}  & \frac{x_9^*}{\sqrt{2}}\\
     0    & x_6 & -x_5 &0  & x_4  & 0  & 0 &-x_7^* &-\frac{x_0}{\sqrt{2}} & \frac{x_8}{\sqrt{2}}\\
     x_7 &0 &0 & x_4^*  &0 & x_5^*   &-x_7^*  & 0  & \frac{x_0^*}{\sqrt{2}} &-\frac{x_8^*}{\sqrt{2}}\\
     0    & x_7  & 0  & -x_5 & 0 & x_4 & 0 & x_6  &-\frac{x_1}{\sqrt{2}}  & \frac{x_{10}}{\sqrt{2}}\\
     0    & 0 & x_7 &-x_6  & 0 & 0 & x_4 &-x_5^* &-\frac{x_2}{\sqrt{2}}   & \frac{x_{10}}{\sqrt{2}}\\
     0    & 0 & 0 & 0 & x_7 &-x_6 & x_5   & x_4^*  &-\frac{x_3}{\sqrt{2}}  & \frac{x_{11}}{\sqrt{2}} \\
     x_{12} &-x_{13}^* &-x_{14}^* &-x_{15}^*  & 0 & 0 & 0 & 0  &-\frac{x_{11}^*}{\sqrt{2}} &-\frac{x_3^*}{\sqrt{2}}\\
     x_{13}& x_{12}^* & 0 & 0  & -x_{14}^* &-x_{15}^* & 0 & 0  & \frac{x_{10}^*}{\sqrt{2}} & \frac{x_2^*}{\sqrt{2}}\\
    x_{14} &0  & x_{12}^* & 0  & x_{13}^* &0 &-x_{15}^* & 0 &-\frac{x_9^*}{\sqrt{2}} &-\frac{x_1^*}{\sqrt{2}}\\
     0    & x_{14} & -x_{13} &0  & x_{12}  & 0  & 0 &-x_{15}^* &-\frac{x_8}{\sqrt{2}} &-\frac{x_0}{\sqrt{2}}\\
     x_{15} &0 &0 & x_{12}^*  &0 & x_{13}^*   &-x_{15}^*  & 0  & \frac{x_8^*}{\sqrt{2}} & \frac{x_0^*}{\sqrt{2}}\\
     0    & x_{15}  & 0  & -x_{13} & 0 & x_{12} & 0 & x_{14}  &-\frac{x_9}{\sqrt{2}} &-\frac{x_1}{\sqrt{2}}\\
     0    & 0 & x_{15} &-x_{14}  & 0 & 0 & x_{12} &-x_{13}^* &-\frac{x_{10}}{\sqrt{2}}   &-\frac{x_2}{\sqrt{2}}\\
     0    & 0 & 0 & 0 & x_{15} &-x_{14} & x_{13}   & x_{12}^*  &-\frac{x_{11}}{\sqrt{2}}  &-\frac{x_3}{\sqrt{2}}
\end{array}\right].
\end{equation}
}

\subsection{Maximal rate of the scaled CODs}
\label{subsec2-4}
It has been shown by Liang \cite{Lia} that the maximal rate of a COD for $n$ transmit antennas is $\frac{1}{2}+\frac{1}{2t}$ when $n=2t-1$ or $2t$.

The following result says when there is scaling of atleast one column then the maximal rate is $\frac{1}{2}.$
\begin{theorem}
The maximal rate of a scaled COD, with scaling of at least one column, for $n$ transmit antennas is $\frac{1}{2}.$
\end{theorem}
\begin{proof}
Let $(DR)_n$ be a scaled COD for $n$ transmit antennas and there exists at-least one column of the matrix such that whenever a variable appears in that column, it is scaled by $\frac{1}{\sqrt{2}}.$  Since all the variables appearing in a column is scaled by $\frac{1}{\sqrt{2}}$, each variable must appear twice in that column. Let the number of distinct complex variables in $(DR)_n$ is $k$. Then $2k\leq p$, i.e., $k/p\leq 1/2$. 
\end{proof}

Thus if one allows to incorporate a factor of $\frac{1}{\sqrt{2}}$ for all the entries of any column, there won't be any improvement in the rate, on the other hand, as we will have observed, we can construct some rate 1/2 codes with lesser decoding delay.
\section{Delay-minimality for 9 transmit antennas}
\label{sec3}
In this section, it is  shown that the low-delay rate $1/2$ COD developed  in the 
previous section is of minimal delay for 9 transmit antennas. 
To prove this, we need some preliminary facts regarding the interrelationship between real and complex ODs and certain bilinear maps. The ROD and bilinear maps are intimately related in the sense that a ROD of size $[p,n,k]$ exists if and only if there exists a normed bilinear map of the same size. The normed bilinear maps have been studied extensively and one find a good introduction to this topic in the book by Shapiro \cite{Sha}. As the results from the theory of normed bilinear maps is used to prove our claim, these maps have been defined below and some facts are stated regarding these maps.

A bilinear map $f$ (over a field $\mathbb{F}$) is a map
\begin{eqnarray}
       f : \mathbb{F}^k \times \mathbb{F}^n &\rightarrow& \mathbb{F}^p \\
             (x,y) &\mapsto&    f(x,y)
\end{eqnarray}
\noindent
such that it is linear in both $x$ and $y$, i.e., $f(x_1+x_2,y)=f(x_1,y)+f(x_2,y)$ and $f(x,y_1+y_2)=f(x,y_1)+f(x,y_2)$ for all $x,x_1,x_2\in\mathbb{F}^k$ and $y,y_1,y_2\in\mathbb{F}^n.$ The space $\mathbb{F}^p$ is called the target space of $f$.
If the vector spaces under consideration are inner product spaces, for example, when the field is real numbers or complex numbers, the Euclidean norm of a vector $x$ is denoted by $\parallel x\parallel$.
If a bilinear map preserves the norm, then it is called a  normed bilinear map. More precisely,
\begin{defn}
 A  {\textit{normed real bilinear map}} (NRBM) of size $[p,n,k]$ is a map
 $f : \mathbb{R}^k \times \mathbb{R}^n \rightarrow \mathbb{R}^p$ such that
 $f$ is bilinear and normed i.e., $\parallel f(x,y)\parallel =\parallel  x\parallel  \parallel y\parallel \forall x\in\mathbb{R}^k, y\in\mathbb{R}^n$. If $f(x,y)=0$ implies $x=0$ or $y=0$, then such a map is called a nonsingular map.
\end{defn}

The following theorem gives a lower bound on $p$ for fixed values of $n$ and $k$.
 \begin{theorem}[Hopf-Stiefel Theorem~\cite{Sha}]
If there exists a nonsingular bilinear map of size $[p,n,k]$ over $\mathbb{R}$, then
      $(x+y)^p =0 \mbox{ in the ring }\mathbb{F}_2[x,y]/(x^n,y^k)$.
\end{theorem}
\begin{defn}
 Let $n,k$ be positive integers. Then the three quantities $n\circ k,$  $p_{BL}$ and $p_{NBL}$ are defined by
\begin{itemize}
\item $n\circ k  = min \{ p: (x+y)^p = 0 \ \mbox {in} \ {\mathbb{F}_2} [x,y]/(x^n,y^k)\}$,\item    $p_{BL}(n,k) = min \{p: $ there is a nonsingular bilinear map $ [p,n,k] $ over $\mathbb{R}  $ \},
\item   $p_{NBL}(n,k) = min \{p: $ there is a normed bilinear map $ [p,n,k]$ over $ \mathbb{R} $\},
\end{itemize}
\end{defn}
The following  basic facts about these quantities are well known~\cite{Sha}.\\
 $p_{NBL}(n,k) \geq p_{BL}(n,k) \geq n\circ k$ and $p_{NBL}(n,k) = n$ if and only if $k \leq \rho (n)$ where $\rho$ is the Hurwitz-Radon function.

It follows from the definition of $ n \circ k $ that
   \begin{prpn}[\cite{Sha}]
   \label{ncirck}
      $ n \circ k $ is a commutative binary operation. \\ 
    $(I)$ If $ k \leq l $ then $ n \circ k \leq  n \circ l $ \\
    $(II)$ $ n \circ k  = 2^m $   if and only if $ k,n \leq 2^m $ and $ k + n > 2^m $ .\\
    $(III)$ If $ n \leq 2^m $ then $ n \circ ( k  + 2^m ) = n \circ k + 2^m . $
\end{prpn}
\begin{example}
 Let us compute $10 \circ 10$.
  We observe that $10 < 2^4$, but $(10 +10) > 16$.
So, $10\circ 10 =16$.
\end{example}

The relation between RODs and NRBMs has been observed by Wang and Xia in \cite{WaX}.
For the sake of completeness and since the proof of this fact gives the explicit relation between the NRBM and the row-vector representation matrices of the ROD which is in correspondence with it, we give here the proof of the following lemma.
\begin{lemma}
\label{rod_nrbl}
A ROD of size $[p,n,k]$ exists if and only if there exists a normed real bilinear map of size $[p,n,k]$.
\end{lemma}
\begin{proof}
Let $\underline{x}= (x_1,x_2,...,x_k)^\mathcal{T},$ 
$\underline{y}=(y_1,\cdots,y_n)^\mathcal{T}$ and
$\underline{z}=(z_1,\cdots,,z_p)^\mathcal{T}$ be real column vectors. 

Let $A$ be a ROD of size $[p,n,k]$ given by $ A=\sum_{i=1}^{k} A_ix_i $ where the $p\times n$ real matrices are the dispersion matrices or weight matrices defining the design $A$ \cite{Lia}. Let 
\begin{eqnarray*}
f : \mathbb{R}^k \times \mathbb{R}^n &\rightarrow& \mathbb{R}^p\\
(\underline{x},\underline{y})&\mapsto&   (\sum_{i=1}^{k} A_i x_i )\underline{y}.
\end{eqnarray*}
We show that $f$ is a normed real bilinear map of size $[p,n,k]$.

Let $\underline{z}=f(\underline{x},\underline{y})$. Then $z_i=\underline{x}^\mathcal{T}B_i\underline{y}$ where the $k\times n$ real matrices $B_i, ~~i=1,2,\cdots,p,$ are  the row vector representation \cite{Lia} of the design $A.$ This representation shows that $f$ is bilinear. Moreover, $f$ is normed, since $\parallel f(\underline{x},\underline{y})\parallel^2=\parallel A\underline{y}\parallel^2=(A\underline{y})^\mathcal{T}A\underline{y}$=
$\underline{y}^\mathcal{T}(x_1^2 + x_2^2 +...+x_k^2)I_n)\underline{y}=\parallel \underline{x}\parallel^2 \parallel \underline{y}\parallel^2$.

To show that the converse holds, let $f$ be the normed bilinear map given by
\begin{eqnarray*}
f : \mathbb{R}^k \times \mathbb{R}^n &\rightarrow& \mathbb{R}^p\\
                                (\underline{x},\underline{y})&\mapsto&   \underline{z}.
\end{eqnarray*}
As $f$ is linear in both $\underline{x}$ and $\underline{y}$, we have $\underline{z}=A\underline{y}$ where $A$ is an $p\times n$ matrix where each entry of the matrix is a real linear combination of the variables $x_1,\cdots,x_k$.
As $f$ is normed, we have
$\parallel \underline{z}\parallel^2=\parallel f(\underline{x},\underline{y})\parallel^2=\parallel \underline{x}\parallel^2 \parallel\underline{y}\parallel^2$.
But $f(\underline{x},\underline{y})=A\underline{y}$.
Then, $\parallel A\underline{y}\parallel^2=((x_1^2 +\cdots+x_k^2)I_n)\underline{y}^\mathcal{T}\underline{y}$.
So, we have $$\underline{y}^\mathcal{T}A\underline{y}=((x_1^2 +\cdots+x_k^2)I_n)\underline{y}^\mathcal{T}\underline{y}.$$
As $\underline{y}$ consists of variables, the above equation is equivalent to $ A^\mathcal{T}A=(x_1^2 + x_2^2 +...+x_k^2)I_n$.
 \end{proof}
We now prove the main result of this section.
\begin{theorem}
The minimum decoding  delay of the rate-$1/2$ COD admitting linear combination of complex variables for $9$ transmit antennas is $16$.
\end{theorem}
\begin{proof}
 In this proof, we assume that the ROD and COD admit linear combination of two or more variables as its entries. 
Suppose, there exists a COD  of size $[2x,9,x]$ where $x$ is an integer less than $8$. This implies existence of a ROD of size $[4x,18,2x], x< 8$ which is obtained by replacing each complex entry by its $2\times 2$ real matrix representation. In the remaining part of the proof, we show that such a ROD does not exist thus proving the theorem.\\
If a  ROD of size $[4x,18,2x]$ exists, then there also exists a normed bilinear map of the same size by Lemma \ref{rod_nrbl} which implies that $4x\geq 18 \circ 2x$.
As $18 \circ 2x\geq 18$, we have $4x\geq 18$ i.e., $x\geq 5$ for $x$ being an integer.
Thus we have three possible choices for $x$, namely $5,6$ and $7$.\\
By Proposition \ref{ncirck}, we have
$18 \circ 2x=26,28,30$ for $x=5,6$ and $7$ respectively.
In all the cases, $18 \circ 2x> 4x$ which contradicts the fact that $4x\geq 18 \circ 2x$.
\end{proof}

\section{PAPR reduction of rate-1/2 scaled CODs}
\label{sec4}
In this section, we study the PAPR properties of Scaled CODs constructed in this paper. Note that in the construction of $TJC_n$ given in \cite{TJC}, even though the delay is more, there are no zero entries in the design matrix. On the contrary, in our construction of low-day codes $(DR)_n$ there are zero entries. To be specific, observe that the first eight columns of rate-$1/2$ code $(DR)_n, n\geq 9$ given in \eqref{gn} contains as many zero as the number of non-zero entries in it, while there is no zero in the remaining columns of the matrix. When the  number of transmit antennas $n$ is more than $7$, the total number of zeros in the codeword matrix is equal to $8(\nu(n)/2)=4\nu(n)$.  Hence the fraction of zeros in the codeword matrix is equal to $\frac{4\nu(n)}{n\nu(n)}=4/n$ for $n\geq 8$.

Now in the remaining part of this section, we show that one can further reduce the number of zeros in $(DR)_n$ by suitably choosing a post-multiplication matrix to it without increasing signaling complexity \cite{DaR3} of the code. 

As seen easily, only  the first eight column  contain zeros while the others do not. Moreover, the zeros in $0$-th column and the $7-$th column occupy complementary locations, so is also for the pairs of columns given by $(1,6), (2,5)$ and $(3,4)$. What it essentially suggests is that we can perform some elementary column operations which will result in a code in which all the entries are non-zero.
 In other words, if the rate-1/2 COD is of size $p\times n$, then we post-multiply it with a matrix $Q_n$ of size $n\times n$ given by
\begin{eqnarray*}
\label{matrix8}
Q_n=\left[\begin{array}{cc}
A & 0  \\
0 & I_{n-8}
\end{array}\right]
\end{eqnarray*}
where $A$  is a matrix of size $8\times 8$ given by
\begin{eqnarray*}
A=\frac{1}{\sqrt{2}}\left[\begin{array}{r@{\hspace{0.9pt}}r@{\hspace{0.9pt}}r@{\hspace{0.9pt}}r@{\hspace{0.9pt}}r@{\hspace{0.9pt}}r@{\hspace{0.9pt}}r@{\hspace{0.9pt}}r@{\hspace{0.9pt}}}
  1&0&0&0&0&0&0&1 \\
  0&1&0&0&0&0&1&0  \\
  0&0&1&0&0&1&0&0  \\
  0&0&0&1&1&0&0&0  \\
  0&0&0&1&-&0&0&0 \\
  0&0&1&0&0&-&0&0 \\
  0&1&0&0&0&0&-&0 \\
  1&0&0&0&0&0&0&-
\end{array}\right]
\end{eqnarray*}
and $I_{n-8}$ is the $(n-8)\times (n-8)$ identity matrix,  then all the entries of the scaled COD given by $(DR)_nQ_n,$ are non-zero. We formally present this fact as:
\begin{theorem}
The matrix $Q_n$ when post-multiplied with a rate-1/2 scaled COD $(DR)_n$ given by
\begin{equation}
   (DR)_n=\left[\begin{array}{cc}
     E_8 & H_t\\
     O_8 & \hat{H}_t
\end{array}\right].
\end{equation}
always gives a COD with no zeros. Moreover, the matrix $Q_n$ does not depend on any particular construction procedure (namely the maps $\gamma_t$ and $\psi_t$) used to obtain the constituent rate-1 RODs.
\end{theorem}
\begin{proof}
It is clear that the first $8$ columns of the matrix has $50\%$ zeros in it and in the remaining $n-8$ columns formed by $H_t$ and $\hat{H}_t$, there are no zeros  as both these matrices are constructed from rate-1 ROD by substituting all the variables in it with appropriate $8$-tuple column vectors. Here neither rate-1 ROD nor the $8$-tuple column vector has any any zero in it. Therefore, the matrix $Q_n$ gives a rate $1/2$ scales COD without any zeros irrespective of how the rate-1 RODs are obtained for the construction of $(DR)_n$.
\end{proof}
\begin{example}
The rate-1/2 code with no zero entry for $9$ transmit antennas is given by

{\footnotesize
\begin{equation*}
  \left[\begin{array}{ccccccccc}
  y_0 &  -y_1^*&  -y_2^*&  -y_3^*   &y_3^*& -y_2^*& -y_1^* &  y_0   & -y_7^*\\
  y_1   &  y_0^*   & -y_3^*   & -y_2^*   & -y_2^*   &  y_3^*   &  y_0^*   & y_1   &  y_6^*\\
  y_2   & -y_3^*   &  y_0^*   &  y_1^*   &  y_1^*   &  y_0^*   &  y_3^*   & y_2   & -y_5^*\\
 -y_3^*   & y_2   &-y_1   & y_0   & y_0   &-y_1   & y_2   &  y_3^*   &-y_4\\
  y_3   &  y_2^*   &  y_1^*   &  y_0^*   & -y_0^*   & -y_1^*   & -y_2^*   & y_3   &  y_4^*\\
  y_2^*   & y_3   & y_0   &-y_1   & y_1   &-y_0   & y_3   & -y_2^*   &-y_5\\
 -y_1^*   & y_0   & y_3   &-y_2   & y_2   & y_3   &-y_0   &  y_1^*   &-y_6\\
  y_0^*   & y_1   &-y_2   & y_3   & y_3   & y_2   &-y_1   & -y_0^*   &-y_7\\
  y_4   & -y_5^*   & -y_6^*   & -y_7^*   & -y_7^*   & -y_6^*   & -y_5^*   & y_4   & -y_3^*\\
  y_5   &  y_4^*   & -y_7^*   & -y_6^*   &  y_6^*   &  y_7^*   &  y_4^*   & y_5   &  y_2^*\\
  y_6   & -y_7^*   &  y_4^*   &  y_5^*   & -y_5^*   &  y_4^*   &  y_7^*   & y_6   & -y_1^*\\
 -y_7^*   & y_6   &-y_5   & y_4   &-y_4   &-y_5   & y_6   &  y_7^*   &-y_0\\
  y_7   &  y_6^*   &  y_5^*   &  y_4^*   &  y_4^*   & -y_5^*   & -y_6^*   & y_7   &  y_0^*\\
  y_6^*   & y_7   & y_4   &-y_5   &-y_5   &-y_4   & y_7   & -y_6^*   &-y_1\\
 -y_5^*   & y_4   & y_7   &-y_6   &-y_6   & y_7   &-y_4   &  y_5^*   &-y_2\\
  y_4^*   & y_5   &-y_6   & y_7   &-y_7   & y_6   &-y_5   & -y_4^*   &-y_3
\end{array}\right]
\end{equation*}
}
\end{example}
\section{Discussion}
\label{sec5}

This paper gives rate-$1/2$ CODs for $n$ transmit antennas with decoding delay equal to $\nu(n)$. The decoding delay of these codes is half of that of the rate-1/2 CODs given in 
Tarokh et al \cite{TJC}. As the maximal rate of a scaled COD is close to $1/2$ for large  number of transmit antennas, the codes constructed in this paper are better than the codes constructed by Liang \cite{Lia} or Lu et al \cite{LFX} when considered for large number of transmit antennas. Another advantage with the designs reported in this paper is that they do not contain zero entries leading to low PAPR.

All the four constructions namely Adams, Lax and Phillips construction from Quaternions, Octonion, Geramita-Pullman construction and the construction given in this paper will give the same square ROD if number of transmit antennas is less than or equal to $8$. Therefore, these four construction will generate the same rate $1/2$ scaled COD if the number of transmit antennas ( of the scaled COD) is less than or equal to $16$. For more than $16$ antennas, rate-1/2 scaled CODs will vary with the methods chosen for the  construction of rate-1 RODs.
Due to the largeness of the matrices involved, it is not possible to display two distinct rate-1/2 scaled CODs for $17$ transmit antennas, obtained by two different construction procedures for rate-1 RODs.

It is not known whether the low-delay  for rate $1/2$ scaled CODs we have achieved is minimal delay except for the case of $9$ transmit antennas. We conjecture that $\nu(n)$ is the minimum value of the decoding delay of rate-1/2 scaled CODs for any $n$ transmit antennas. It will be interesting to see whether this is indeed true.

An interesting direction for further research would be to investigate whether the necessary conditions given in Theorem \ref{ratesquarerod} on the maps $\gamma_t$ and $\psi_t$ are indeed sufficient also. 


\begin{appendices}
\section{Recursive Construction of $R_t$}
\label{appendixI}
In this appendix we show that the RODs $R_t$ can be constructed recursively. 

Let $K_t=B_t$ for $t=1,2,4$ and $8$.
The four square ODs $K_t,t=1,2,4,8$ are shown below.
{\scriptsize
\begin{eqnarray}
\label{K1248}
(x_0),~
\left(\begin{array}{rr}
    x_0   & x_1\\
   -x_1   & x_0
\end{array}\right), ~
\left(\begin{array}{rrrr}
    x_0   & x_1   & x_2    & x_3 \\
   -x_1   & x_0   &-x_3    & x_2 \\
   -x_2   & x_3   & x_0    &-x_1 \\
   -x_3   &-x_2   & x_1    & x_0 \nonumber\\
\end{array}\right),\\
\left(\begin{array}{rrrrrrrr}
    x_0   & x_1   & x_2    & x_3    & x_4    & x_5     & x_6  & x_7\\
   -x_1   & x_0   &-x_3    & x_2    &-x_5    & x_4     & x_7  &-x_6 \\
   -x_2   & x_3   & x_0    &-x_1    &-x_6    &-x_7     & x_4  & x_5 \\
   -x_3   &-x_2   & x_1    & x_0    &-x_7    & x_6     &-x_5  & x_4 \\
   -x_4   & x_5   & x_6    & x_7    & x_0    &-x_1     &-x_2  &-x_3 \\
   -x_5   &-x_4   & x_7    &-x_6    & x_1    & x_0     & x_3  &-x_2 \\
   -x_6   &-x_7   &-x_4    & x_5    & x_2    &-x_3     & x_0  & x_1 \\
   -x_7   & x_6   &-x_5    &-x_4    & x_3    & x_2     &-x_1  & x_0
\end{array}\right).
\end{eqnarray}
}
It follows that
\begin{eqnarray*}
\begin{array}{c}
K_t^\mathcal{T}=K_t^\mathcal{T}(x_0,x_1,\cdots,x_{t-1})=K_t(x_0,-x_1,\cdots,-x_{t-1})\\\mbox{ and }-K_t^\mathcal{T}=K_t(-x_0,x_1,\cdots,x_{t-1})
\end{array}
\end{eqnarray*}
for $t=1,2,4$ or $8$.
The expression for $R_t$  of order $t$ as given in Theorem \ref{orthog} gives rise to the following recursive construction of $R_t$.
\begin{figure*}
\begin{equation*}
  R_{2n}= \left[\begin{array}{cc}
 R_n & x_{\rho(n)}I_n \\
-x_{\rho(n)}I_n & R_n^\mathcal{T}
\end{array}\right],
\quad
  R_{4n}= \left[\begin{array}{cc}
 R_{2n} & x_{\rho(n)+1}I_{2n}\\
-x_{\rho(n)+1}I_{2n} & R_{2n}^\mathcal{T}
\end{array}\right],
  \end{equation*}
 \begin{equation}
\label{Rn16n1}
  R_{8n}= \left[\begin{array}{cc}
 R_{4n} & T_4(y_0,y_1)\otimes I_n \\
T_4(-y_0,y_1)\otimes I_n & R_{4n}^\mathcal{T}
\end{array}\right],
\quad
 R_{16n}= \left[\begin{array}{cc}
 R_{8n} & T_8(y_2,y_3,y_4,y_5)\otimes I_n \\
 T_8(-y_2,y_3,y_4,y_5)\otimes I_n & R_{8n}^\mathcal{T}
\end{array}\right]\\
\end{equation}
\hrule
\end{figure*}
Given two matrices $U=(u_{ij})$ of size $v_1\times w_1$ and $V$ of size $v_2\times w_2$, we define the {\it{Kronecker product or tensor product }} of $U$ and $V$ as the following $v_1v_2\times w_1w_2$ matrix:
\[\left(\begin{array}{cccc}
  u_{11}V     &u_{12}V         & \cdots &u_{1w_1}V  \\
  u_{11}V     &u_{12}V         & \cdots &u_{1w_1}V  \\
  \vdots      &\vdots          & \ddots &\vdots     \\
  u_{v_11}V   &u_{v_12}V       & \cdots &u_{v_1w_1}V
\end{array}\right).  \]

Let $I_n$ be an identity matrix of size $n$. Define
\begin{eqnarray*}
I_2^0=\begin{bmatrix} 1 & 0 \\ 0 & 1 \end{bmatrix},&&~
I_2^1=\begin{bmatrix} 1 & 0 \\ 0 & -1 \end{bmatrix},\\
I_2^2=\begin{bmatrix} 0 & 1 \\ 1 & 0 \end{bmatrix},&& ~
I_2^3=\begin{bmatrix} 0 & -1 \\ 1 & 0 \end{bmatrix},\\
I_4^0=I_4,&& I_4^1=I_2^3 \otimes I_2^2,\\
I_8^0=I_8,&& I_8^1=I_2^0 \otimes I_4^1,\\
I_8^2= I_2^3 \otimes I_2^1 \otimes I_2^2,&& I_8^3= I_2^3 \otimes I_2^2 \otimes I_2^0.
\end{eqnarray*}

Let $y_0,\cdots,y_5$ be real variables. Define
\begin{eqnarray*}
T_4(y_0,y_1)&=&y_0I_4^0 +y_1I_4^1,\\
T_8(y_2,y_3,y_4,y_5)&=& y_2 I_8^0 + y_3 I_8^1 +y_4 I_8^2 + y_5 I_8^3.
\end{eqnarray*}

We have four RODs of order $n=2^a$ with $a=0,1,2,3$ as given in
\eqref{K1248} which are respectively $K_1,K_2,K_4$ and $K_8$.\\
Assuming that a square ROD of order $n=2^{4l-1}, l\geq 1$
\[
R_n=R_n(x_0,\cdots,x_{\rho(n)-1})
\]
which has $\rho(n)$ real variables, is given, then
we construct $R_{2n},R_{4n},R_{8n},R_{16n}$ of order $2n$, $4n$, $8n$ and $16n$ respectively given by \eqref{Rn16n1}, as shown at the top of the next page
 where $y_i=x_{\rho(n)+2+i}$ and
\begin{eqnarray*}
R_t^\mathcal{T}&=&R_t^\mathcal{T}(x_0,x_1,\cdots,x_{\rho(t)-1})\\
               &=&R_t(x_0,-x_1,\cdots,-x_{\rho(t)-1}),\\
-R_t^\mathcal{T}&=&R_t(-x_0,x_1,\cdots,x_{\rho(t)-1}).
\end{eqnarray*}
 
\section{Adams-Lax-Phillips and Geramita-Pullman constructions as special cases}
\label{appendixII}
In this appendix we show that the well known constructions of square RODs by Adams-Lax-Phillips using Octonions and Quaternions as well as the construction by Geramita and Pullman are nothing but our construction corresponding to specific choices of the functions  $\gamma_t$ and $\psi_t$ defined by \eqref{gammamap} and \eqref{psimap}. 
It turns out to be convenient to use the map  $\chi_t=\psi_t\gamma_t$ instead of the map $\psi_t.$ Note that both $\gamma_t$ and $\chi_t$ act on the set $Z_{\rho(t)}$ and are injective. Now given $\gamma_t$ and $\chi_t$, we have $\psi_t=\chi_t\gamma_t^{(-1)}$.
With this new definition, we can reformulate the criteria given in Theorem \ref{orthog}
as follows.
\begin{eqnarray}
\label{gammachi}
\vert (\chi_t(x)\oplus \chi_t(y))\cdot(\gamma_t(x)\oplus \gamma_t(y))\vert \\
\mbox{ is an odd integer } \forall x,y\in Z_{\rho(t)}, x\neq y \nonumber.
\end{eqnarray}
In the following lemma, we define $\gamma_t$ and $\chi_t$ in three different ways and these maps are shown to satisfy the relation given in \eqref{gammachi}.
Although both $\gamma_t$ and $\chi_t$ are different for all the three cases for arbitrary values of $t$, $\gamma_t$ is the identity map when $t=1,2,4$ or $8$.
Hence $\chi_t=\psi_t$ if $t\in\{1,2,4,8\}$ and is given by \eqref{psi1248}.

\begin{lemma}
\label{threedesign}
Let $t=2^a$, $a=4c+d$, $m\in\{0,1,\cdots,7\}$.
Let $\gamma_t$ and $\chi_t$ be two maps defined over $Z_{\rho(t)}$ in three different ways as given below. Identify $\gamma_t(Z_{\rho(t)})$ and $\chi_t(Z_{\rho(t)})$ as subsets of $\mathbb{F}_2^a$.
Then $\vert (\gamma_t(x_1)\oplus \gamma_t(x_2))\cdot (\chi_t(x_1)\oplus \chi_t(x_2))\vert$ is odd for all $x_1,x_2\in Z_{\rho(t)}$, $x_1\neq x_2$.\\
For $x=8l+m \in Z_{\rho(t)},$

(i)
\begin{eqnarray*}
\gamma_t(8l+m)&=&t(1-2^{-l})+{8^l}m,\nonumber\\
\chi_t(8l+m)&=&
\begin{cases}
0  &  \text{ if  $l=0, m=0$} \\
t.2^{-l} & \text{ if  $l\neq 0, m=0$} \\
8^l\chi_{2^d}(m)&\text{ if $l=c, m\neq 0$ }\\
t.2^{-l-1}+8^l\chi_8(m)&\text{ if $l\neq c, m\neq 0$, }
 \end{cases}
\end{eqnarray*}

(ii)
{\footnotesize
\begin{eqnarray*}
\gamma_t(8l+m)&=&
 \begin{cases}
t(1-2^{-2l})+2^{2l}m  & \hspace{-8pt}\mbox{ if }  0\leq m \leq 3  \\
t(1-2^{-2l-1})+2^{2l}(m-4)  & \hspace{-8pt}\mbox{ if } 4\leq m \leq 7,  \nonumber\\
\end{cases}\\
\chi_t(8l+m)&=&
\begin{cases}
0  &  \mbox{ if } l=0,m=0 \\
t.2^{-2l} & \text{ if  $l\neq 0,m=0$}\\
t.2^{-2l-1}  &  \text{ if  $l\neq 0,m=4$} \\
4            &  \text{ if  $l=0,m=4$} \\
2^{2l}\chi_{2^d}(m)&\text{ if $l=c,m\neq 0$ }\\
t.2^{-2l-1}+2^{2l}\chi_4(m)&\text{ if $l\neq c,m\in\{1,2,3\}$ }\\
t.2^{-2l-2}+2^{2l}\chi_4^\prime(m-4)&\text{ if $l\neq c,m\in\{5,6,7\}$ },\\
 \end{cases}
\end{eqnarray*}
}
where $\chi_4^\prime=\left(\begin{array}{cccccccccccc}
    0   & 1 &2 &3 \\
    0   & 1 &3 &2
    \end{array}\right).$

(iii)
\begin{eqnarray*}
\gamma_t(8l+m)&=&
 \begin{cases}
\frac{8t}{15}(1-2^{-4l}) +\frac{tm}{16^{l+1}} &\text{ if $l< c$, }\\
\frac{8t}{15}(1-2^{-4l}) +m & \text{ if $l=c$ }\nonumber\\
\end{cases}\\
\chi_t(8l+m)&=&
\begin{cases}
0  &  \text{ if  $l=0,m=0$} \\
\frac{t}{2}2^{-4(l-1)} & \text{ if  $l\neq 0, m=0$}\\
 \chi_{2^d}(m)&\text{ if $l=c, m\neq 0$. }\\
\frac{t}{2}2^{-4l}+\frac{t\chi_8(m)}{2^{4(l+1)}}&\text{ if $l\neq c,m\neq 0$. }
 \end{cases}
\end{eqnarray*}
\end{lemma}
\begin{proof}
We give proof only for the case (i). The cases  (ii) and (iii) can be proved similarly.

It is enough to prove that \\
  (B1) $\vert \gamma_t(x)\cdot \chi_t(x)\vert$ is odd for all $x\neq 0$, $x\in Z_{\rho(t)}$ and \\
  (B2) $\vert \gamma_t(x_1)\cdot \chi_t(x_2)\vert +\vert\gamma_t(x_2)\cdot \chi_t(x_1)\vert$ is odd for all $x_1,x_2\in Z_{\rho(t)}$,  $x_1\neq x_2, x_1\neq 0,x_2\neq 0$.

Let $\gamma_t(8l+m)=\gamma_t^{(1)}(8l+m)+\gamma_t^{(2)}(8l+m)$ such that $\gamma_t^{(1)}(8l+m)=t(1-2^{-l})$ and $\gamma_t^{(2)}(8l+m)={8^l}m$.\\
Similarly, let $\chi_t(8l+m)=\chi_t^{(1)}(8l+m)+\chi_t^{(2)}(8l+m)$ such that 
\begin{eqnarray}
\chi_t^{(1)}(8l+m)=
    \begin{cases}
0  &  \text{ if  $l=0,m=0,$} \nonumber\\
t2^{-l} & \text{ if  $l\neq 0,m=0,$} \nonumber\\
0 &\text{ if $l=c,m \neq 0,$ }\\
t2^{-l-1} &\text{ if $l\neq c,m\neq 0,$ }\\
 \end{cases}\\
\chi_t^{(2)}(8l+m)=
    \begin{cases}
0  &  \text{ if  $l=0,m=0,$} \nonumber\\
0  & \text{ if  $l\neq 0,m=0,$} \nonumber\\
8^l\chi_{2^d}(m)&\text{ if $l=c,m\neq 0,$ }\\
8^l\chi_{8}(m)&\text{ if $l\neq c,m\neq 0.$ }
 \end{cases}
\end{eqnarray}

\noindent Let $8l+m\neq 0$ and $8l^\prime+ m^\prime\neq 0$.
From the definition of $\gamma_t^i,\chi_t^i,i=1,2$, it follows that\\
\begin{eqnarray*}
\begin{array}{l}
(A1)~~ \vert\chi_t^{(2)}(8l+m)\cdot \gamma_t^{(2)}(8l^\prime+ m^\prime)\vert=0 \mbox{ if } l\neq l^\prime,\\
(A2)~~ \vert\chi_t^{(1)}(8l+m)\cdot \gamma_t^{(1)}(8l^\prime+ m^\prime)\vert=1\mbox{ if }l <l^\prime,\\
(A3)~~ \vert\chi_t^{(1)}(8l+m)\cdot \gamma_t^{(1)}(8l^\prime+ m^\prime)\vert=0\mbox{ if }l >l^\prime \\
\hspace{24pt}\mbox{ or if } l=l^\prime, m\neq 0, \\
(A4)~~ \vert\chi_t^{(1)}(8l)\cdot \gamma_t^{(1)}(8l+m)\vert=1 \mbox{ if }l\neq 0, \\
(A5)~~ \vert\chi_t^{(1)}(x)\cdot \gamma_t^{(2)}(y)\vert=\vert\chi_t^{(2)}(x)\cdot \gamma_t^{(1)}(y)\vert=0\\
     \forall~ x,y\in Z_{\rho(t)},\\
(A6)~~ \vert\chi_t^{(2)}(8l)\cdot \gamma_t^{(2)}(8l+m)\vert=\vert\chi_t^{(2)}(8l+m)\cdot \gamma_t^{(2)}(8l)\vert=0.
\end{array}
\end{eqnarray*}

First we prove (B1). Let $x=8l+m$ with $m\neq 0$. We have 

{\footnotesize
\begin{eqnarray*}
\vert \chi_t(x)\cdot\gamma_t(x)\vert&\equiv&\vert\chi_t^{(1)}(8l+m)\cdot \gamma_t^{(1)}(8l+m)\vert + \vert\chi_t^{(2)}(8l+m)\\
&& \cdot \gamma_t^{(2)}(8l+m)\vert + \vert\chi_t^{(1)}(8l+m)\cdot \gamma_t^{(2)}(8l+m)\vert \\
&& + \vert\chi_t^{(2)}(8l+m)\cdot \gamma_t^{(1)}(8l+m)\vert\\
&=&\vert\chi_t^{(1)}(8l+m)\cdot \gamma_t^{(1)}(8l+m)\vert\\ &&+\vert\chi_t^{(2)}(8l+m)\cdot \gamma_t^{(2)}(8l+m)\vert \mbox{ by (A5) } \\
 &=&\vert\chi_t^{(2)}(8l+m)\cdot \gamma_t^{(2)}(8l+m)\vert \mbox{ using (A3) }\\	
&=&\vert\chi_e(m)\cdot m\vert, \mbox{ $e=2^d$ if $l=c$, else $e=8$ }
\end{eqnarray*}
}
\noindent
But $\vert\chi_e(m)\cdot m\vert$ is an odd number by Lemma \ref{propertyV8}.\\
If $m=0$, we have $\vert \gamma_t(x)\cdot \chi_t(x)\vert=1$ by (A4).\\

To prove (B2), let $x_1\neq 0$ and $x_2\neq 0$.
Write $x_2=8l_2+m_2$, $x_1=8l_1+m_1$ with $x_2>x_1$. 
We have two cases:\\
(C1): $l_2>l_1$,~~~~~~
(C2): $l_2=l_1=l$, $m_2> m_1$.\\

\noindent
{\bf Case (C1):} we have 

{\footnotesize
\begin{eqnarray*}
\chi_t(x_2)\cdot \gamma_t(x_1)=\chi_t^{(1)}(8l_2+m_2)\cdot \gamma_t^{(1)}(8l_1+m_1) \\
\vspace{120pt}\oplus \chi_t^{(2)}(8l_2+m_2)\cdot \gamma_t^{(2)}(8l_1+m_1) \text{ by (A5) }.
\end{eqnarray*}
}
\noindent
But $\vert\chi_t^{(1)}(8l_2+m_2)\cdot \gamma_t^{(1)}(8l_1+m_1)\vert=0$ by (A3) \\
and $\vert\chi_t^{(2)}(8l_2+m_2)\cdot \gamma_t^{(2)}(8l_1+m_1)\vert=0$ by (A1),\\
thus $\vert\chi_t(x_2)\cdot \gamma_t(x_1)\vert=0$. \\
Now $\chi_t(x_1)\cdot \gamma_t(x_2)= \chi_t^{(1)}(8l_1+m_1)\cdot \gamma_t^{(1)}(8l_2+m_2) \oplus \chi_t^{(2)}(8l_1+m_1)\cdot \gamma_t^{(2)}(8l_2+m_2)$ by (A5).\\
But $\vert\chi_t^{(2)}(8l_1+m_1)\cdot \gamma_t^{(2)}(8l_2+m_2)\vert=0$ by (A1) and
$\vert \chi_t^{(1)}(8l_1+m_1)\cdot \gamma_t^{(1)}(8l_2+m_2)\vert=1$ by (A2).\\
Hence $\vert \chi_t(x_1)\cdot \gamma_t(x_2)\vert + \vert \chi_t(x_2)\cdot \gamma_t(x_1)\vert$
is an odd number.

\noindent
{\textbf{Case (C2):}} we consider two following cases:\\
 (i) $m_1\neq 0$ and (ii) $m_1=0$. Note that $m_2$ is always non-zero.\\
Let $d=\vert (\chi_t(x_1)\cdot \gamma_t(x_2))\oplus(\chi_t(x_2)\cdot \gamma_t(x_1))\vert$.\\

\noindent
{\textbf{Case (i):}} We have
 {\footnotesize
\begin{eqnarray*}
 d &\equiv&\vert\chi_t^{(2)}(8l+m_1)\cdot\gamma_t^{(2)}(8l+m_2)\vert \\
&&+ \vert \chi_t^{(2)}(8l+m_2)\cdot \gamma_t^{(2)}(8l+m_1)\vert \mbox{ by (A3) and (A5) }\\
&=&\vert(\chi_e(m_1)\cdot m_2)
\oplus(\chi_e(m_2)\cdot m_1)\vert, \mbox{ $e=2^d$ if $l=c$, else $e=8$ }
\end{eqnarray*}
}
which is an odd number by Lemma \ref{propertyV8}.\\
{\textbf{Case (ii):}}
Since $m_1=0$, therefore $l\neq 0$. We have
\begin{eqnarray*}
 d &\equiv&\vert\chi_t^{(1)}(8l)\cdot\gamma_t^{(2)}(8l+m_2)\vert \\
&&+\vert \chi_t^{(1)}(8l+m_2)\cdot \gamma_t^{(1)}(8l)\vert \mbox{ by (A6). }\\
&=& 1 \mbox{ by (A3) and (A4). }
\end{eqnarray*}

\end{proof}
\begin{figure*}
{\small
\begin{equation}
\label{ALPQ}
  \mathbb{O}_3= \left(\begin{array}{cccc}
I_n \otimes L_4(x_0,x_1,x_2,x_3)& 0_{4n} & I_n \otimes R_4(x_4,x_5,x_6,x_7) & \mathbb{O}_1(y_0,\cdots,y_{\rho(n)-1})\otimes I_4 \\
  0_{4n} & I_n \otimes L_4(x_0,x_1,x_2,x_3) & -\mathbb{O}_1^\mathcal{T}(y_0,\cdots,y_{\rho(n)-1})\otimes I_4  & I_n \otimes R_4^\mathcal{T}(x_4,x_5,x_6,x_7) \\
 I_n \otimes -R_4^\mathcal{T}(x_4,x_5,x_6,x_7) & \mathbb{O}_1(y_0,\cdots,y_{\rho(n)-1})\otimes I_4 & I_n \otimes L_4^\mathcal{T}(x_0,x_1,x_2,x_3)
& 0_{4n} \\
-\mathbb{O}_1^\mathcal{T}(y_0,\cdots,y_{\rho(n)-1})\otimes I_4 & I_n \otimes -R_4(x_4,x_5,x_6,x_7) &  0_{4n} & I_n\otimes L_4^\mathcal{T}(x_0,x_1,x_2,x_3)
\end{array}\right)
\end{equation}
}
\hrule
\end{figure*}
By Lemma \ref{threedesign} and Theorem \ref{ratesquarerod}, the matrix $B_t$ defined by the two functions $\gamma_t$ and $\chi_t$ is a square ROD in all the three cases. We refer to these three different RODs by $A_t, \hat{A}_t$ and $P_t$ corresponding to the pair of functions defined in (i), (ii) and (iii) respectively. Observe that our  construction $R_t$ is different from any of $A_t, \hat{A}_t$ and $P_t$ for general values of $t.$

Now, we proceed to show that the designs $A_t, \hat{A}_t$ and $P_t$ are essentially the Adams-Lax-Phillips construction using Octonions and Quaternions and the Geramita-Pullman construction respectively with change in sign of some rows or columns. 
 
\subsection{Adams-Lax-Phillips Construction from Octonions as a special case}
The Adams-Lax-Phillips construction from Octonions is given by induction from order $n=2^a$ to $16n$ as follows \cite{Lia}:
Denoting the square ROD of order $n=2^a$ resulting from the Adams-Lax-Phillips construction using Octonions by  
\[
\mathbb{O}_n=\mathbb{O}_n(x_0,\cdots,x_{\rho(n)-1})
\]
which has $\rho(n)$ real variables, the square ROD of order $16n$ with $(\rho(n)+8)$
real variables $x_i,$ $i=0,1,\cdots, \rho(n)+7,$
\[
\mathbb{O}_{16n}=\mathbb{O}_{16n}(x_0,\cdots,x_{\rho(n)+7})
\]
is given by
\begin{eqnarray*}
  \mathbb{O}_{16n}= \left[\begin{array}{cc}
 I_n\otimes K_8(y_0,\cdots,y_7) & \mathbb{O}_n\otimes I_8 \\
\mathbb{O}_n^\mathcal{T}\otimes I_8 & I_n\otimes (-K_8^\mathcal{T}(y_0,\cdots,y_7))\\
\end{array}\right]
 \end{eqnarray*}
with $y_i=x_{\rho(n)+i}$. \\
With re-arrangement of variables and change in signs, we rewrite the design $\mathbb{O}_{16n}$ as

{\footnotesize
\begin{eqnarray}
\label{ALPO}
  \mathbb{O}_{16n}^{(O)}= \left[\begin{array}{cc}
 I_n\otimes K_8(x_0,\cdots,x_7) & \mathbb{O}_n^{(O)}(y_0,\cdots,y_{\rho(n)-1})\otimes I_8 \\
-\mathbb{O}_n^{(O)\mathcal{T}}(y_0,\cdots,y_{\rho(n)-1})\otimes I_8 & I_n\otimes K_8^\mathcal{T}(x_0,\cdots,x_7)\\
\end{array}\right]
 \end{eqnarray}
}
with $y_i=x_{8+i}$ and $\mathbb{O}_n^{(O)}=\mathbb{O}_n,~~ n=1,2,4,8.$ The reason why we consider this rearranged version is that we show in Lemma \ref{AtO2} that $A_t$ is same as $\mathbb{O}_{2n}^{(O)}$ with $t=16n.$\\
\begin{lemma}
\label{AtO2}
Let $t \geq 16$ and a power of $2.$ Also, let $A_t$ be the square ROD of order $t$ as given in Lemma \ref{threedesign} (i), and $\mathbb{O}_{16n}^{(O)}$ be the square ROD given in \eqref{ALPO} which is of order $16n$. Then $A_t=\mathbb{O}_{16n}^{(O)}$ for $t=16n$.
\end{lemma}
\begin{proof} We prove it by induction on $t$.
For $t=1,2,4$ and $8$, $A_t=K_t$ and the COD $\mathbb{O}_t^{(O)}$ of order $t$ is also given by $K_t$. Hence the lemma holds for $t=1,2,4$ and $8$.
Assuming that the lemma holds for $t=n$, i.e., $A_n=\mathbb{O}_n^{(O)}$ of order $n$, we have to prove that the lemma also holds for $t=16n, $ i.e., $A_{16n}=\mathbb{O}_{16n}^{(O)}.$

Let
\begin{equation}
  A_{16n}= \left[\begin{array}{cc}
 \hat{A}_{11} & \hat{A}_{12} \\
\hat{A}_{21} & \hat{A}_{22}
\end{array}\right]
 \end{equation}
where $\hat{A}_{ij}$, $1\leq i,j\leq 2$ are square matrices of size $8n\times 8n$.
It is easy to check that the location of non-zero variables in the matrix $A_{16n}$ coincide with that of $\mathbb{O}_{16n}^{(O)}$. Therefore it is enough to show the signs (positive/negative polarity) of the corresponding entries in the two designs are same i.e.,
\begin{enumerate}
\item $\mu_{16n}(i,j)= \mu_{16n}(i\% 8, j\% 8 )$ for $0\leq i,j \leq 8n-1$,\\
\item $\mu_{16n}(i,j)= \mu_8(i,j)$ for $0\leq i,j \leq 7$,\\
\item $\mu_{16n}(i,j)=\mu_{16n}(i\oplus i\%8, j\oplus j\%8)$ \\
if $0\leq i\leq 8n-1, 8n\leq j \leq 16n-1$,\\
\item $\mu_{16n}(8i,8n\oplus 8j)=\mu_n(i,j)$ for $0\leq i,j \leq n-1$, \\
\item $\mu_{16n}(8n \oplus i,8n \oplus j)=\mu_{16n}(i,j)$ if $i\oplus j=0$ or $i\oplus j>8n$,\\
\item $\mu_{16n}(8n \oplus i,8n \oplus j)=-\mu_{16n}(i,j)$ \\
if $i\oplus j\in\{1,2,\cdots,7\}\cup \{8n\}$.\\ 
\end{enumerate}

Note that\\
1) \& 2) together imply $\hat{A}_{11}=I_n\otimes K_8(x_0,\cdots,x_7 )$,\\
3) \& 4) together imply $\hat{A}_{12}=\mathbb{O}_{n}^{(O)}\otimes I_8$ and \\
5) \& 6) together imply $\hat{A}_{22}=A_{11}^\mathcal{T}, \hat{A}_{21}=-A_{12}^\mathcal{T}$.\\
\noindent
Let $A_{16n}(i,j)\neq 0$.\\
Then $i\oplus j\in \hat{Z}_{\rho(16n)}$ and $\mu_{16n}(i,j)=(-1)^{\vert i\cdot\psi_{16n}(i\oplus j)\vert}$.\\
To prove 1), we have to show that
$\vert i\cdot\psi_{16n}(i\oplus j)\vert\equiv \vert (i\% 8)\cdot\psi_{16n}(i\% 8\oplus j\% 8)$
for $0\leq i,j \leq 8n-1$.\\
We have $i\oplus j=(16n)(1-2^{-l})+{8^l}m$ and $i\oplus j< 8n$. So $l=0$ and $i\oplus j=m$. i.e., $i \oplus j=i\% 8 \oplus j\%8$.\\
Thus it is enough to prove that $\vert (i\oplus i\% 8)\cdot\psi_{16n}(i\oplus j)\vert\equiv 0$
Now
$(i\oplus i\% 8)< 8n$, $8$ divides $(i\oplus i\% 8)$ and $\psi_{16n}(i\oplus j)=8n\oplus\psi_8(m)$, hence the statement holds.

The statement 2) is true as
$\vert i\cdot\psi_{16n}(i\oplus j)\vert\equiv \vert i\cdot\psi_8(i\oplus j)\vert$
for $0\leq i,j \leq 7$. \\ 
In order to prove 3), we must have

{\small
\begin{eqnarray*}
 \vert i\cdot\psi_{16n}(i\oplus j)\vert &\equiv& \vert (i\oplus i\% 8)\cdot\psi_{16n}((i \oplus i\% 8) \oplus (j\oplus j\% 8))\vert
\end{eqnarray*}
}

i.e., $\vert (i\% 8)\cdot\psi_{16n}((i \oplus i\% 8) \oplus (j\oplus j\% 8))\vert\equiv 0$.
As $8n\leq i\oplus j\leq 16n-1$, we have $i\oplus j=(16n)(1-2^{-l})+{8^l}m$ with $l\geq 1$.
So $8$ divides $i \oplus j$ as $8$ divides both $(16n)(1-2^{-l})$ and ${8^l}m$.
So $i\% 8=j\% 8$ i.e., $i\oplus j=((i \oplus i\%8) \oplus (j\oplus j\%8))$.
Thus it is enough to prove that $\vert (i\% 8)\cdot\psi_{16n}(i \oplus j)\vert \equiv 0$.
It is indeed true as $\psi_{16n}(i \oplus j)$ is a multiple of $8$.

To prove 4), we have to show that
\[
\vert (8i)\cdot\psi_{16n}(8n\oplus 8i\oplus 8j)\vert \equiv \vert (i\cdot\psi_{n}((i \oplus j).
\]
We have $8n\oplus 8i\oplus 8j=(16n)(1-2^{-l})+8^lm$ for some $l$ with $l\geq 1$ and $m\in Z_8$.
Let $16n=2^a$ and $a=4c+d$.\\
If $l=c$, we have $\psi_{16n}(8n\oplus 8i\oplus 8j)=8^l\chi_{2^d}(m)$ and $\psi_{n}(i\oplus j)=8^{l-1}\chi_{2^d}(m)$. One can easily see that the above statement holds.\\
On the other hand, if $l<c$, we have  $\psi_{16n}(8n\oplus 8i\oplus 8j)=(16n)2^{-l-1}+8^l\chi_{8}(m)$ and $\psi_{n}(i\oplus j)=n.2^{-l}+8^{l-1}\chi_{8}(m)$.
In this case too, the statement holds.

To prove 5), we have to show that
\[ \vert (i\oplus 8n)\cdot\psi_{16n}(i\oplus j)\vert \equiv \vert i\cdot\psi_{16n}(i\oplus j)\vert,
\]
i.e., $\vert (8n)\cdot\psi_{16n}(i\oplus j)\vert\equiv 0$.
Now for $i\oplus j=0$ or greater than $8n$, $ (8n)\cdot\psi_{16n}(i\oplus j)=\mathbf{0}$.

To prove 6), we have to show that
\[ \vert (i\oplus 8n)\cdot\psi_{16n}(i\oplus j)\vert \equiv 1+\vert i\cdot\psi_{16n}(i\oplus j)\vert,
\]
i.e., $\vert (8n)\cdot\psi_{16n}(i\oplus j)\vert\equiv 1$.
But $(8n)\cdot\psi_{16n}(i\oplus j)=8n$ for all $(i\oplus j)\in\{1,2,3,4,5,6,7,8n\}$.
\end{proof}
\subsection{Adams-Lax-Phillips Construction from Quaternions and Geramita-Pullman Construction as special cases}
Adams-Lax-Phillips has also provided another construction of square RODs using Quaternions which is explicitly shown in \cite{Lia}. 

Assuming that a square ROD of order $n=2^a$
\[
\mathbb{O}_n^{(Q)}=\mathbb{O}_n^{(Q)}(x_0,\cdots,x_{\rho(n)-1})
\]
which has $\rho(n)$ real variables, is given, then a square ROD of order $16n$ with $\rho(n)+8$
real variables $x_i$ for  $i=0,1,\cdots, \rho(n)+7$
\[
\mathbb{O}_{16n}^{(Q)}=\mathbb{O}_{16n}^{(Q)}(x_0,\cdots,x_{\rho(n)+7})
\]
is given by \eqref{ALPQ}, as shown at the top of this page where
the two matrices $L_4$ and $R_4$ are given by
\begin{equation*}
L_4(x_0,x_1,x_2,x_3)=\left(\begin{array}{rrrr}
    x_0   & x_1   & x_2    & x_3 \\
   -x_1   & x_0   &-x_3    & x_2 \\
   -x_2   & x_3   & x_0    &-x_1 \\
   -x_3   &-x_2   & x_1    & x_0
\end{array}\right),
\end{equation*}

\begin{equation*}
R_4(x_4,x_5,x_6,x_7)=\left(\begin{array}{rrrr}
    x_4   & x_5   & x_6    & x_7 \\
   -x_5   & x_4   & x_7    &-x_6 \\
   -x_6   &-x_7   & x_4    & x_5 \\
   -x_7   & x_6   &-x_5    & x_4
\end{array}\right).
\end{equation*}
respectively with $y_i=x_{8+i}$.\\
The Geramita-Pullman construction of square ROD is also given by induction explicitly in \cite{Lia}.

Consider a recursive construction of square ROD of order $n=2^a$ to $16n$ as follows:
\[
\mathbb{O}_n^{(GP)}=\mathbb{O}_n^{(GP)}(x_0,\cdots,x_{\rho(n)-1})
\]
which has $\rho(n)$ real variables is given, then a square ROD $ \mathbb{O}_{16n}^{(GP)}$ of order $16n$ with $\rho(n)+8$
real variables $x_i$ for  $i=0,1,\cdots, \rho(n)+7$ is given by

{\footnotesize
\begin{eqnarray}
\label{geramita}
 \left[\begin{array}{ll}
  K_8(x_0,\cdots,x_7)\otimes I_n & I_8 \otimes\mathbb{O}_n^{(GP)}(y_0,\cdots,y_{\rho(n)-1})\\
I_8\otimes (-\mathbb{O}_n^{(GP)})^\mathcal{T}(y_0,\cdots,y_{\rho(n)-1}) & K_8^\mathcal{T}(x_0,\cdots,x_7)\otimes I_n\\
\end{array}\right]
 \end{eqnarray}
}
with $y_i=x_{8+i}$. \\
It can be checked  that both Adams-Lax-Phillips construction from Quaternions and Geramita-Pullman's construction given in \cite{Lia} differ from the constructions of $\mathbb{O}_{16n}^{(Q)}$ and $\mathbb{O}_{16n}^{(GP)}$ defined above only in rearrangement of variables and in signs of some of the rows or columns of the design matrix.
\begin{lemma}
\label{AtQGP}
Let $t \geq 16$ and $\hat{A}_t$ and $P_t$ be the square ROD of order $t$ as given in Lemma \ref{threedesign} (ii) and (iii), and also let $\mathbb{O}_{16n}^{(Q)}$ and $\mathbb{O}_{16n}^{(GP)}$ be the square RODs given in \eqref{ALPQ} and in \eqref{geramita} which are of order $16n$. Then $\hat{A}_t=\mathbb{O}_{16n}^{(Q)}$ and $P_t=\mathbb{O}_{16n}^{(GP)}$ for $t=16n.$   
\end{lemma}
\begin{proof}
Similar to that of Lemma \ref{AtO2} and hence omitted.
\end{proof}
\begin{example}
\label{ADPGPexamples}
Square ROD $A_{16}$ of size $[16,16,9]$ by Adams-Lax-Phillips construction from Octonion is given by \eqref{ALPOc}.

\begin{figure*}
\begin{eqnarray}
\label{ALPOc}
\left[\begin{array}{rrrrrrrrrrrrrrrr}
  x_0 &    x_1 &    x_2 &    x_3 &    x_4 &    x_5 &    x_6 &    x_7 &    x_8 &     0 & 0 &     0 &     0 &     0 &     0 &     0\\
 -x_1 &    x_0 &   -x_3 &    x_2 &   -x_5 &    x_4 &    x_7 &   -x_6 &     0 &    x_8 & 0 &     0 &     0 &     0 &     0 &     0\\
 -x_2 &    x_3 &    x_0 &   -x_1 &   -x_6 &   -x_7 &    x_4 &    x_5 &     0 &     0 &    x_8 &     0 &     0 &     0 &     0 &     0\\
 -x_3 &   -x_2 &    x_1 &    x_0 &   -x_7 &    x_6 &   -x_5 &    x_4 &     0 &     0 &     0 &    x_8 &     0 &     0 &     0 &     0\\
 -x_4 &    x_5 &    x_6 &    x_7 &    x_0 &   -x_1 &   -x_2 &   -x_3 &     0 &     0 &     0 &     0 &    x_8 &     0 &     0 &     0\\
 -x_5 &   -x_4 &    x_7 &   -x_6 &    x_1 &    x_0 &    x_3 &   -x_2 &     0 &     0 &     0 &     0 &     0 &    x_8 &     0 &     0\\
 -x_6 &   -x_7 &   -x_4 &    x_5 &    x_2 &   -x_3 &    x_0 &    x_1 &     0 &     0 &     0 &     0 &     0 &     0 &    x_8 &     0\\
 -x_7 &    x_6 &   -x_5 &   -x_4 &    x_3 &    x_2 &   -x_1 &    x_0 &     0 &     0 &     0 &     0 &     0 &     0 &     0 &    x_8\\
  x_8 &     0 &     0 &     0 &     0 &     0 &     0 &     0 &   -x_0 &    x_1 &    x_2 &   x_3 &    x_4 &    x_5 &    x_6 &    x_7\\
   0 &    x_8 &     0 &     0 &     0 &     0 &     0 &     0 &   -x_1 &   -x_0 &   -x_3 &   x_2 &   -x_5 &    x_4 &    x_7 &   -x_6\\
   0 &  0 & x_8 &     0 &     0 &     0 &     0 &     0 &   -x_2 &    x_3 &   -x_0 &   -x_1 & -x_6 & -x_7 &    x_4 &    x_5\\
   0 &     0 &     0 &    x_8 &     0 &     0 &     0 &     0 &   -x_3 &   -x_2 &    x_1 &  -x_0 &   -x_7 &    x_6 &   -x_5 &    x_4\\
   0 &     0 &     0 &     0 &    x_8 &     0 &     0 &     0 &   -x_4 &    x_5 &    x_6 &   x_7 &   -x_0 &   -x_1 &   -x_2 &   -x_3\\
   0 &     0 &     0 &     0 &     0 &    x_8 &     0 &     0 &   -x_5 &   -x_4 &    x_7 &  -x_6 &    x_1 &   -x_0 &    x_3 &   -x_2\\
   0 &     0 &     0 &     0 &     0 &     0 &    x_8 &     0 &   -x_6 &   -x_7 &   -x_4 &   x_5 &    x_2 &   -x_3 &   -x_0 &    x_1\\
   0 &     0 &     0 &     0 &     0 &     0 &     0 &    x_8 &   -x_7 &    x_6 &   -x_5 &  -x_4 &    x_3 &    x_2 &   -x_1 &   -x_0\\
\end{array}\right]
\end{eqnarray}
\hrule
\end{figure*}
Square ROD $\hat{A}_{16}$ of size $[16,16,9]$ by Adams-Lax-Phillips construction from Quaternion is given by \eqref{ALPQu}.
\begin{figure*}
\begin{eqnarray}
\label{ALPQu}
\left[\begin{array}{rrrrrrrrrrrrrrrr}
  x_0 &    x_1 &    x_2 &    x_3 &     0 &     0 &     0 &     0 &    x_4 &    x_5 &    x_6 &    x_7 &    x_8 &     0 &     0 &     0\\
 -x_1 &    x_0 &   -x_3 &    x_2 &     0 &     0 &     0 &     0 &   -x_5 &    x_4 &   -x_7 &   -x_6 &     0 &    x_8 &     0 &     0\\
 -x_2 &    x_3 &    x_0 &   -x_1 &     0 &     0 &     0 &     0 &   -x_6 &   -x_7 &    x_4 &    x_5 &     0 &     0 &    x_8 &     0\\
 -x_3 &   -x_2 &    x_1 &    x_0 &     0 &     0 &     0 &     0 &   -x_7 &    x_6 &   -x_5 &    x_4 &     0 &     0 &     0 &    x_8\\
   0 &     0 &     0 &     0 &    x_0 &    x_1 &    x_2 &    x_3 &   -x_8 &     0 &     0 &     0 &    x_4 &   -x_5 &   -x_6 &   -x_7\\
   0 &     0 &     0 &     0 &   -x_1 &    x_0 &   -x_3 &    x_2 &     0 &   -x_8 &     0 &     0 &    x_5 &    x_4 &   -x_7 &    x_6\\
   0 &     0 &     0 &     0 &   -x_2 &    x_3 &    x_0 &   -x_1 &     0 &     0 &   -x_8 &     0 &    x_6 &   -x_7 &    x_4 &   -x_5\\
   0 &     0 &     0 &     0 &   -x_3 &   -x_2 &    x_1 &    x_0 &     0 &     0 &     0 &  -x_8 &    x_7 &   -x_6 &    x_5 &    x_4\\
 -x_4 &    x_5 &    x_6 &    x_7 &    x_8 &     0 &     0 &     0 &    x_0 &   -x_1 &   -x_2 &   -x_3 &     0 &     0 &     0 &     0\\
 -x_5 &   -x_4 &    x_7 &   -x_6 &     0 &    x_8 &     0 &     0 &    x_1 &    x_0 &    x_3 &   -x_2 &     0 &     0 &     0 &     0\\
 -x_6 &    x_7 &   -x_4 &    x_5 &     0 &     0 &    x_8 &     0 &    x_2 &   -x_3 &    x_0 &    x_1 &     0 &     0 &     0 &     0\\
 -x_7 &    x_6 &   -x_5 &   -x_4 &     0 &     0 &     0 &    x_8 &    x_3 &    x_2 &   -x_1 &    x_0 &     0 &     0 &     0 &     0\\
 -x_8 &     0 &     0 &     0 &   -x_4 &   -x_5 &   -x_6 &   -x_7 &     0 &     0 &     0 &     0 &    x_0 &   -x_1 &   -x_2 &   -x_3\\
   0 &   -x_8 &     0 &     0 &    x_5 &   -x_4 &    x_7 &    x_6 &     0 &     0 &     0 &     0 &    x_1 &    x_0 &    x_3 &   -x_2\\
   0 &     0 &   -x_8 &     0 &    x_6 &    x_7 &   -x_4 &   -x_5 &     0 &     0 &     0 &     0 &    x_2 &   -x_3 &    x_0 &    x_1\\
   0 &     0 &     0 &   -x_8 &    x_7 &   -x_6 &    x_5 &   -x_4 &     0 &     0 &     0 &     0 &    x_3 &    x_2 &   -x_1 &    x_0\\
\end{array}\right]
\end{eqnarray}
\hrule
\end{figure*}
Square ROD of $P_{16}$ size $[16,16,9]$ by Geramita-Pullman construction is the same as $R_{16}.$

Square ROD of $P_{32}$ size $[32,32,10]$ by Geramita-Pullman construction is given by \eqref{GP32}.

\vspace*{5.00cm}
\begin{figure*}
\tiny{
\begin{eqnarray}
\label{GP32}
\left[\begin{array}{ r @{\hspace{.2pt}} r @{\hspace{.2pt}} r @{\hspace{.2pt}} r @{\hspace{.2pt}}r @{\hspace{.2pt}} r @{\hspace{.2pt}} r @{\hspace{.2pt}} r @{\hspace{.2pt}}
r @{\hspace{.2pt}} r @{\hspace{.2pt}} r @{\hspace{.2pt}} r @{\hspace{.2pt}}r @{\hspace{.2pt}} r @{\hspace{.2pt}} r @{\hspace{.2pt}} r @{\hspace{.2pt}} r @{\hspace{.2pt}} r @{\hspace{.2pt}} r @{\hspace{.2pt}} r @{\hspace{.2pt}}r @{\hspace{.2pt}} r @{\hspace{.2pt}} r @{\hspace{.2pt}} r @{\hspace{.2pt}}r @{\hspace{.2pt}} r @{\hspace{.2pt}} r @{\hspace{.2pt}} r @{\hspace{.2pt}}r @{\hspace{.2pt}} r @{\hspace{.2pt}} r @{\hspace{.2pt}} r @{\hspace{.2pt}}}
  x_0 &     0 &    x_1 &     0 &    x_2 &     0 &    x_3 &     0 &    x_4 &     0 &    x_5 &     0 &    x_6 &     0 &    x_7 &     0 &    x_8 &    x_9 &     0 &     0 &     0 &     0 &     0 &     0 &     0 &     0 &     0 &     0 &     0 &     0 &     0 &     0\\
   0 &    x_0 &     0 &    x_1 &     0 &    x_2 &     0 &    x_3 &     0 &    x_4 &     0 &    x_5 &     0 &    x_6 &     0 &    x_7 &   -x_9 &    x_8 &     0 &     0 &     0 &     0 &     0 &     0 &     0 &     0 &     0 &     0 &     0 &     0 &     0 &     0\\
 -x_1 &     0 &    x_0 &     0 &   -x_3 &     0 &    x_2 &     0 &   -x_5 &     0 &    x_4 &     0 &    x_7 &     0 &   -x_6 &     0 &     0 &     0 &    x_8 &    x_9 &     0 &     0 &     0 &     0 &     0 &     0 &     0 &     0 &     0 &     0 &     0 &     0\\
   0 &   -x_1 &     0 &    x_0 &     0 &   -x_3 &     0 &    x_2 &     0 &   -x_5 &     0 &    x_4 &     0 &    x_7 &     0 &   -x_6 &     0 &     0 &   -x_9 &    x_8 &     0 &     0 &     0 &     0 &     0 &     0 &     0 &     0 &     0 &     0 &     0 &     0\\
 -x_2 &     0 &    x_3 &     0 &    x_0 &     0 &   -x_1 &     0 &   -x_6 &     0 &   -x_7 &     0 &    x_4 &     0 &    x_5 &     0 &     0 &     0 &     0 &     0 &    x_8 &    x_9 &     0 &     0 &     0 &     0 &     0 &     0 &     0 &     0 &     0 &     0\\
   0 &   -x_2 &     0 &    x_3 &     0 &    x_0 &     0 &   -x_1 &     0 &   -x_6 &     0 &   -x_7 &     0 &    x_4 &     0 &    x_5 &     0 &     0 &     0 &     0 &   -x_9 &    x_8 &     0 &     0 &     0 &     0 &     0 &     0 &     0 &     0 &     0 &     0\\
 -x_3 &     0 &   -x_2 &     0 &    x_1 &     0 &    x_0 &     0 &   -x_7 &     0 &    x_6 &     0 &   -x_5 &     0 &    x_4 &     0 &     0 &     0 &     0 &     0 &     0 &     0 &   x_8 &    x_9 &     0 &     0 &     0 &     0 &     0 &     0 &     0 &     0\\
   0 &   -x_3 &     0 &   -x_2 &     0 &    x_1 &     0 &    x_0 &     0 &   -x_7 &     0 &    x_6 &     0 &   -x_5 &     0 &    x_4 &     0 &     0 &     0 &     0 &     0 &     0 &  -x_9 &    x_8 &     0 &     0 &     0 &     0 &     0 &     0 &     0 &     0\\
 -x_4 &     0 &    x_5 &     0 &    x_6 &     0 &    x_7 &     0 &    x_0 &     0 &   -x_1 &     0 &   -x_2 &     0 &   -x_3 &     0 &     0 &     0 &     0 &     0 &     0 &     0 &    0 &     0 &    x_8 &    x_9 &     0 &     0 &     0 &     0 &     0 &     0\\
   0 &   -x_4 &     0 &    x_5 &     0 &    x_6 &     0 &    x_7 &     0 &    x_0 &     0 &   -x_1 &     0 &   -x_2 &     0 &   -x_3 &     0 &     0 &     0 &     0 &     0 &     0 &    0 &     0 &   -x_9 &    x_8 &     0 &     0 &     0 &     0 &     0 &     0\\
 -x_5 &     0 &   -x_4 &     0 &    x_7 &     0 &   -x_6 &     0 &    x_1 &     0 &    x_0 &     0 &    x_3 &     0 &   -x_2 &     0 &     0 &     0 &     0 &     0 &     0 &     0 &    0 &     0 &     0 &     0 &    x_8 &    x_9 &     0 &     0 &     0 &     0\\
   0 &   -x_5 &     0 &   -x_4 &     0 &    x_7 &     0 &   -x_6 &     0 &    x_1 &     0 &    x_0 &     0 &    x_3 &     0 &   -x_2 &     0 &     0 &     0 &     0 &     0 &     0 &    0 &     0 &     0 &     0 &   -x_9 &    x_8 &     0 &     0 &     0 &     0\\
 -x_6 &     0 &   -x_7 &     0 &   -x_4 &     0 &    x_5 &     0 &    x_2 &     0 &   -x_3 &     0 &    x_0 &     0 &    x_1 &     0 &     0 &     0 &     0 &     0 &     0 &     0 &    0 &     0 &     0 &     0 &     0 &     0 &    x_8 &    x_9 &     0 &     0\\
   0 &   -x_6 &     0 &   -x_7 &     0 &   -x_4 &     0 &    x_5 &     0 &    x_2 &     0 &   -x_3 &     0 &    x_0 &     0 &    x_1 &     0 &     0 &     0 &     0 &     0 &     0 &    0 &     0 &     0 &     0 &     0 &     0 &   -x_9 &    x_8 &     0 &     0\\
 -x_7 &     0 &    x_6 &     0 &   -x_5 &     0 &   -x_4 &     0 &    x_3 &     0 &    x_2 &     0 &   -x_1 &     0 &    x_0 &     0 &     0 &     0 &     0 &     0 &     0 &     0 &    0 &     0 &     0 &     0 &     0 &     0 &     0 &     0 &    x_8 &    x_9\\
   0 &   -x_7 &     0 &    x_6 &     0 &   -x_5 &     0 &   -x_4 &     0 &    x_3 &     0 &    x_2 &     0 &   -x_1 &     0 &    x_0 &     0 &     0 &     0 &     0 &     0 &     0 &    0 &     0 &     0 &     0 &     0 &     0 &     0 &     0 &   -x_9 &    x_8\\
 -x_8 &    x_9 &     0 &     0 &     0 &     0 &     0 &     0 &     0 &     0 &     0 &  0 &     0 &     0 &     0 &     0 &    x_0 &     0 &   -x_1 &     0 &   -x_2 &     0 &   -x_3 &     0 &   -x_4 &     0 &   -x_5 &     0 &   -x_6 &     0 &   -x_7 &     0\\
 -x_9 &   -x_8 &     0 &     0 &     0 &     0 &     0 &     0 &     0 &     0 &     0 &  0 &     0 &     0 &     0 &     0 &     0 &    x_0 &     0 &   -x_1 &     0 &   -x_2 & 0 &   -x_3 &     0 &   -x_4 &     0 &   -x_5 &     0 &   -x_6 &     0 &   -x_7\\
   0 &     0 &   -x_8 &    x_9 &     0 &     0 &     0 &     0 &     0 &     0 &     0 &  0 &     0 &     0 &     0 &     0 &    x_1 &     0 &    x_0 &     0 &    x_3 &     0 &   -x_2 &     0 &    x_5 &     0 &   -x_4 &     0 &   -x_7 &     0 &    x_6 &     0\\
   0 &     0 &   -x_9 &   -x_8 &     0 &     0 &     0 &     0 &     0 &     0 &     0 &  0 &     0 &     0 &     0 &     0 &     0 &    x_1 &     0 &    x_0 &     0 &    x_3 & 0 &   -x_2 &     0 &    x_5 &     0 &   -x_4 &     0 &   -x_7 &     0 &    x_6\\
   0 &     0 &     0 &     0 &   -x_8 &    x_9 &     0 &     0 &     0 &     0 &     0 &  0 &     0 &     0 &     0 &     0 &    x_2 &     0 &   -x_3 &     0 &    x_0 &     0 &    x_1 &     0 &    x_6 &     0 &    x_7 &     0 &   -x_4 &     0 &   -x_5 &     0\\
   0 &     0 &     0 &     0 &   -x_9 &   -x_8 &     0 &     0 &     0 &     0 &     0 &  0 &     0 &     0 &     0 &     0 &     0 &    x_2 &     0 &   -x_3 &     0 &    x_0 & 0 &    x_1 &     0 &    x_6 &     0 &    x_7 &     0 &   -x_4 &     0 &   -x_5\\
   0 &     0 &     0 &     0 &     0 &     0 &   -x_8 &    x_9 &     0 &     0 &     0 &  0 &     0 &     0 &     0 &     0 &    x_3 &     0 &    x_2 &     0 &   -x_1 &     0 &    x_0 &     0 &    x_7 &     0 &   -x_6 &     0 &    x_5 &     0 &   -x_4 &     0\\
   0 &     0 &     0 &     0 &     0 &     0 &   -x_9 &   -x_8 &     0 &     0 &     0 &  0 &     0 &     0 &     0 &     0 &     0 &    x_3 &     0 &    x_2 &     0 &   -x_1 & 0 &    x_0 &     0 &    x_7 &     0 &   -x_6 &     0 &    x_5 &     0 &   -x_4\\
   0 &     0 &     0 &     0 &     0 &     0 &     0 &     0 &   -x_8 &    x_9 &     0 &  0 &     0 &     0 &     0 &     0 &    x_4 &     0 &   -x_5 &     0 &   -x_6 &     0 &   -x_7 &     0 &    x_0 &     0 &    x_1 &     0 &    x_2 &     0 &    x_3 &     0\\
   0 &     0 &     0 &     0 &     0 &     0 &     0 &     0 &   -x_9 &   -x_8 &     0 &  0 &     0 &     0 &     0 &     0 &     0 &    x_4 &     0 &   -x_5 &     0 &   -x_6 & 0 &   -x_7 &     0 &    x_0 &     0 &    x_1 &     0 &    x_2 &     0 &    x_3\\
   0 &     0 &     0 &     0 &     0 &     0 &     0 &     0 &     0 &     0 &   -x_8 &    x_9 &     0 &     0 &     0 &     0 &    x_5 &     0 &    x_4 &     0 &   -x_7 &     0 &    x_6 &     0 &   -x_1 &     0 &    x_0 &     0 &   -x_3 &     0 &    x_2 &     0\\
   0 &     0 &     0 &     0 &     0 &     0 &     0 &     0 &     0 &     0 &   -x_9 &   -x_8 &     0 &     0 &     0 &     0 &     0 &    x_5 &     0 &    x_4 &     0 &   -x_7 & 0 &    x_6 &     0 &   -x_1 &     0 &    x_0 &     0 &   -x_3 &     0 &    x_2\\
   0 &     0 &     0 &     0 &     0 &     0 &     0 &     0 &     0 &     0 &     0 &     0 &   -x_8 &    x_9 &     0 &     0 &    x_6 &     0 &    x_7 &     0 &    x_4 &     0 &   -x_5 &     0 &   -x_2 &     0 &    x_3 &     0 &    x_0 &     0 &   -x_1 &     0\\
   0 &     0 &     0 &     0 &     0 &     0 &     0 &     0 &     0 &     0 &     0 &     0 &   -x_9 &   -x_8 &     0 &     0 &     0 &    x_6 &     0 &    x_7 &     0 &    x_4 & 0 &   -x_5 &     0 &   -x_2 &     0 &    x_3 &     0 &    x_0 &     0 &   -x_1\\
   0 &     0 &     0 &     0 &     0 &     0 &     0 &     0 &     0 &     0 &     0 &     0 &     0 &     0 &   -x_8 &    x_9 &    x_7 &     0 &   -x_6 &     0 &    x_5 &     0 &    x_4 &     0 &   -x_3 &     0 &   -x_2 &     0 &    x_1 &     0 &    x_0 &     0\\
   0 &     0 &     0 &     0 &     0 &     0 &     0 &     0 &     0 &     0 &     0 &     0 &     0 &     0 &   -x_9 &   -x_8 &     0 &    x_7 &     0 &   -x_6 &     0 &    x_5 & 0 &    x_4 &     0 &   -x_3 &     0 &   -x_2 &     0 &    x_1 &     0 &    x_0\\
\end{array}\right]
\end{eqnarray}
}
\hrule
\end{figure*}
\end{example}
\section{rate 1/2 scaled COD of size $[64,10,32]$}
\label{appendixVI}

{\footnotesize
\begin{equation*}
\hspace{-10pt}
\frac{1}{\sqrt{2}}\cdot
\left[ \hspace{-5pt}
\begin{array}{ r @{\hspace{.2pt}} r @{\hspace{.2pt}} r @{\hspace{.2pt}} r @{\hspace{.2pt}} r @{\hspace{.2pt}} r @{\hspace{.2pt}} r @{\hspace{.2pt}} r @{\hspace{.2pt}} r @{\hspace{.2pt}} r}
    x_0   &-x_1   &-x_2    &-x_3    &-x_4    &-x_5     &-x_6  &-x_7 &-x_8 &-x_{16}\\
    x_1   & x_0   & x_3    &-x_2    & x_5    &-x_4     &-x_7  & x_6 & x_9 & x_{17}\\
    x_2   &-x_3   & x_0    & x_1    & x_6    &x_7      &-x_4  &-x_5 & x_{10}& x_{18}\\
    x_3   & x_2   &-x_1    & x_0    & x_7    &-x_6     & x_5  &-x_4 & x_{11}& x_{19}\\
    x_4   &-x_5   &-x_6    &-x_7    & x_0    & x_1     & x_2  & x_3 & x_{12}& x_{20}\\
    x_5   & x_4   &-x_7    & x_6    &-x_1    & x_0     &-x_3  & x_2 & x_{13}& x_{21}  \\
    x_6   & x_7   & x_4    &-x_5    &-x_2    &x_3      & x_0  &-x_1 & x_{14}& x_{22}  \\
    x_7   &-x_6   & x_5    & x_4    &-x_3    &-x_2     & x_1  & x_0 & x_{15}& x_{23}\\
    x_8&-x_9&-x_{10} &-x_{11} &-x_{12} &-x_{13}  &-x_{14}&-x_{15} & x_0 & x_{24}\\
    x_9& x_8   &-x_{11} & x_{10} &-x_{13} & x_{12}  & x_{15}&-x_{14}&-x_1& x_{25}\\
    x_{10}& x_{11}& x_8    &-x_9 &-x_{14} &-x_{15}  & x_{12}& x_{13}&-x_2& x_{26}\\
    x_{11}&-x_{10}& x_9 & x_8    &-x_{15} & x_{14}  &-x_{13}& x_{12}&-x_3& x_{27}\\
    x_{12}& x_{13}& x_{14} & x_{15} & x_8    &-x_9  &-x_{10}&-x_{11}&-x_4& x_{28}\\
    x_{13}&-x_{12}& x_{15} &-x_{14} & x_9 & x_8     & x_{11}&-x_{10}&-x_5 & x_{29}\\
    x_{14}&-x_{15}&-x_{12} & x_{13} & x_{10} &-x_{11}  & x_8   & x_9&-x_6& x_{30}\\
    x_{15}&x_{14} &-x_{13} &-x_{12} & x_{11} & x_{10}  &-x_9& x_8   &-x_7& x_{31}\\
    x_{16}   &-x_{17}   &-x_{18}    &-x_{19}    &-x_{20}    &-x_{21}     &-x_{22}  &-x_{23} &-x_{24} & x_0\\
    x_{17}   & x_{16}   & x_{19}    &-x_{18}    & x_{21}    &-x_{20}     &-x_{23}  & x_{22} & x_{25}& -x_1\\
    x_{18}   &-x_{19}   & x_{16}    & x_{17}    & x_{22}    &x_{23}      &-x_{20}  &-x_{21} & x_{26}& -x_2\\
    x_{19}   & x_{18}   &-x_{17}    & x_{16}    & x_{23}    &-x_{22}     & x_{21}  &-x_{20} & x_{27}& -x_3\\
    x_{20}   &-x_{21}   &-x_{22}    &-x_{23}    & x_{16}    & x_{17}     & x_{18}  & x_{19} & x_{28}& -x_4\\
    x_{21}   & x_{20}   &-x_{23}    & x_{22}    &-x_{17}    & x_{16}     &-x_{19}  & x_{18} & x_{29} & -x_5 \\
    x_{22}   & x_{23}   & x_{20}    &-x_{21}    &-x_{18}    &x_{19}      & x_{16}  &-x_{17} & x_{30} & -x_6 \\
    x_{23}   &-x_{22}   & x_{21}    & x_{20}    &-x_{19}    &-x_{18}     & x_{17}  & x_{16} & x_{31}& -x_7\\
    x_{24}&-x_{25}&-x_{26} &-x_{27} &-x_{28} &-x_{29}  &-x_{30}&-x_{31} & x_{16}& -x_8\\
    x_{25}& x_{24}   &-x_{27} & x_{26} &-x_{29} & x_{28}  & x_{31}&-x_{30}&-x_{17}& -x_9\\    x_{26}& x_{27}& x_{24}    &-x_{25} &-x_{30} &-x_{31}  & x_{28}& x_{29}&-x_{18}& -x_{10}\\
    x_{27}&-x_{26}& x_{25} & x_{24}    &-x_{31} & x_{30}  &-x_{29}& x_{28}&-x_{19}& -x_{11}\\
    x_{28}& x_{29}& x_{30} & x_{31} & x_{24}    &-x_{25}  &-x_{26}&-x_{27}&-x_{20}& -x_{12}\\
    x_{29}&-x_{28}& x_{31} &-x_{30} & x_{25} & x_{24}     & x_{27}&-x_{26}&-x_{21} & -x_{13}  \\
    x_{30}&-x_{31}&-x_{28} & x_{29} & x_{26} &-x_{27}  & x_{24}   & x_{25}&-x_{22}& -x_{14}\\
    x_{31}&x_{30} &-x_{29} &-x_{28} & x_{27} & x_{26}  &-x_{25}& x_{24}   &-x_{23}& -x_{15}\\
x^*_0   &-x^*_1   &-x^*_2    &-x^*_3    &-x^*_4    &-x^*_5     &-x^*_6  &-x^*_7 &-x^*_8 &-x^*_{16}\\
    x^*_1   & x^*_0   & x^*_3    &-x^*_2    & x^*_5    &-x^*_4     &-x^*_7  & x^*_6 & x^*_9 & x^*_{17}\\
    x^*_2   &-x^*_3   & x^*_0    & x^*_1    & x^*_6    &x^*_7      &-x^*_4  &-x^*_5 & x^*_{10}& x^*_{18}\\
    x^*_3   & x^*_2   &-x^*_1    & x^*_0    & x^*_7    &-x^*_6     & x^*_5  &-x^*_4 & x^*_{11}& x^*_{19}\\
    x^*_4   &-x^*_5   &-x^*_6    &-x^*_7    & x^*_0    & x^*_1     & x^*_2  & x^*_3 & x^*_{12}& x^*_{20}\\
    x^*_5   & x^*_4   &-x^*_7    & x^*_6    &-x^*_1    & x^*_0     &-x^*_3  & x^*_2 & x^*_{13}& x^*_{21}  \\
    x^*_6   & x^*_7   & x^*_4    &-x^*_5    &-x^*_2    &x^*_3      & x^*_0  &-x^*_1 & x^*_{14}& x^*_{22}  \\
    x^*_7   &-x^*_6   & x^*_5    & x^*_4    &-x^*_3    &-x^*_2     & x^*_1  & x^*_0 & x^*_{15}& x^*_{23}\\
    x^*_8&-x^*_9&-x^*_{10} &-x^*_{11} &-x^*_{12} &-x^*_{13}  &-x^*_{14}&-x^*_{15} & x^*_0 & x^*_{24}\\
    x^*_9& x^*_8   &-x^*_{11} & x^*_{10} &-x^*_{13} & x^*_{12}  & x^*_{15}&-x^*_{14}&-x^*_1& x^*_{25}\\
    x^*_{10}& x^*_{11}& x^*_8    &-x^*_9 &-x^*_{14} &-x^*_{15}  & x^*_{12}& x^*_{13}&-x^*_2& x^*_{26}\\
    x^*_{11}&-x^*_{10}& x^*_9 & x^*_8    &-x^*_{15} & x^*_{14}  &-x^*_{13}& x^*_{12}&-x^*_3& x^*_{27}\\
    x^*_{12}& x^*_{13}& x^*_{14} & x^*_{15} & x^*_8    &-x^*_9  &-x^*_{10}&-x^*_{11}&-x^*_4& x^*_{28}\\
    x^*_{13}&-x^*_{12}& x^*_{15} &-x^*_{14} & x^*_9 & x^*_8     & x^*_{11}&-x^*_{10}&-x^*_5 & x^*_{29}\\
    x^*_{14}&-x^*_{15}&-x^*_{12} & x^*_{13} & x^*_{10} &-x^*_{11}  & x^*_8   & x^*_9&-x^*_6& x^*_{30}\\
    x^*_{15}&x^*_{14} &-x^*_{13} &-x^*_{12} & x^*_{11} & x^*_{10}  &-x^*_9& x^*_8   &-x^*_7& x^*_{31}\\
    x^*_{16}   &-x^*_{17}   &-x^*_{18}    &-x^*_{19}    &-x^*_{20}    &-x^*_{21}     &-x^*_{22}  &-x^*_{23} &-x^*_{24} & x^*_0\\
    x^*_{17}   & x^*_{16}   & x^*_{19}    &-x^*_{18}    & x^*_{21}    &-x^*_{20}     &-x^*_{23}  & x^*_{22} & x^*_{25}& -x^*_1\\
    x^*_{18}   &-x^*_{19}   & x^*_{16}    & x^*_{17}    & x^*_{22}    &x^*_{23}      &-x^*_{20}  &-x^*_{21} & x^*_{26}& -x^*_2\\
    x^*_{19}   & x^*_{18}   &-x^*_{17}    & x^*_{16}    & x^*_{23}    &-x^*_{22}     & x^*_{21}  &-x^*_{20} & x^*_{27}& -x^*_3\\
    x^*_{20}   &-x^*_{21}   &-x^*_{22}    &-x^*_{23}    & x^*_{16}    & x^*_{17}     & x^*_{18}  & x^*_{19} & x^*_{28}& -x^*_4\\
    x^*_{21}   & x^*_{20}   &-x^*_{23}    & x^*_{22}    &-x^*_{17}    & x^*_{16}     &-x^*_{19}  & x^*_{18} & x^*_{29} & -x^*_5 \\
    x^*_{22}   & x^*_{23}   & x^*_{20}    &-x^*_{21}    &-x^*_{18}    &x^*_{19}      & x^*_{16}  &-x^*_{17} & x^*_{30} & -x^*_6 \\
    x^*_{23}   &-x^*_{22}   & x^*_{21}    & x^*_{20}    &-x^*_{19}    &-x^*_{18}     & x^*_{17}  & x^*_{16} & x^*_{31}& -x^*_7\\
    x^*_{24}&-x^*_{25}&-x^*_{26} &-x^*_{27} &-x^*_{28} &-x^*_{29}  &-x^*_{30}&-x^*_{31} & x^*_{16}& -x^*_8\\
    x^*_{25}& x^*_{24}   &-x^*_{27} & x^*_{26} &-x^*_{29} & x^*_{28}  & x^*_{31}&-x^*_{30}&-x^*_{17}& -x^*_9\\
    x^*_{26}& x^*_{27}& x^*_{24}    &-x^*_{25} &-x^*_{30} &-x^*_{31}  & x^*_{28}& x^*_{29}&-x^*_{18}& -x^*_{10}\\
    x^*_{27}&-x^*_{26}& x^*_{25} & x^*_{24}    &-x^*_{31} & x^*_{30}  &-x^*_{29}& x^*_{28}&-x^*_{19}& -x^*_{11}\\
    x^*_{28}& x^*_{29}& x^*_{30} & x^*_{31} & x^*_{24}    &-x^*_{25}  &-x^*_{26}&-x^*_{27}&-x^*_{20}& -x^*_{12}\\
    x^*_{29}&-x^*_{28}& x^*_{31} &-x^*_{30} & x^*_{25} & x^*_{24}     & x^*_{27}&-x^*_{26}&-x^*_{21} & -x^*_{13}  \\
    x^*_{30}&-x^*_{31}&-x^*_{28} & x^*_{29} & x^*_{26} &-x^*_{27}  & x^*_{24}   & x^*_{25}&-x^*_{22}& -x^*_{14}\\
    x^*_{31}&x^*_{30} &-x^*_{29} &-x^*_{28} & x^*_{27} & x^*_{26}  &-x^*_{25}& x^*_{24}   &-x^*_{23}& -x^*_{15}
\end{array}\right].
\end{equation*}
}
\end{appendices}
\end{document}